\documentclass[12pt,a4paper]{article}
\usepackage{epsfig}
\usepackage{epic}
\usepackage{float}
\usepackage{afterpage}
\restylefloat{figure}
\renewcommand{\baselinestretch}{1.7}\normalsize

\newcommand{\be}{\begin{equation}}
\newcommand{\ee}{\end{equation}}
\newcommand{\bea}{\begin{eqnarray}}
\newcommand{\eea}{\end{eqnarray}}
\newcommand{\klgl}{\:\hbox to -0.2pt{\lower2.5pt\hbox{$\sim$}\hss}
{\raise3pt\hbox{$<$}}\:}
\newcommand{\grgl}{\:\hbox to -0.2pt{\lower2.5pt\hbox{$\sim$}\hss}
{\raise3pt\hbox{$>$}}\:}

\newcommand{\Gk}{\mbox{$\Gamma_k$}}

\newcommand{\STr}{\mbox{STr}}
\newcommand{\Tr}{\mbox{Tr}}

\begin{document}
%
%
\markboth{ }{ }
\renewcommand{\baselinestretch}{1}\normalsize
\vspace*{-2cm}
\hfill HD-THEP-97-46

\hfill \hfill Revised version, October~17, 1997
\vspace*{2cm}
\bigskip
\bigskip
\begin{center}
{\huge\bf{Effective Quark Interactions and \\ 
QCD-Propagators}} 
\end{center}
\bigskip
\begin{center}
Bastian Bergerhoff\footnote{
B.Bergerhoff@thphys.uni-heidelberg.de; Supported by the Deutsche Forschungsgemeinschaft}$^{,}$\footnote{Address after November $1^{\mbox{st}}$, 1997: Physik-Department, TU M{\"u}nchen, D-85748 Garching} and 
 Christof Wetterich\footnote{
C.Wetterich@thphys.uni-heidelberg.de}\\
Institut f\"ur Theoretische Physik \\
Universit\"at Heidelberg \\
Philosophenweg 16, D-69120 Heidelberg
\end{center}
\setcounter{footnote}{0}
\bigskip
\vspace*{3cm}\begin{abstract}
\noindent
We compute the momentum dependence of the effective four quark interaction in QCD after integrating out the gluons. 
Our method is based on a truncation of exact renormalization group equations which should give reasonable results for momenta above the confinement scale.
The difference between the four quark interaction and the heavy quark potential can be minimized for an optimal renormalization scheme in Landau gauge.
Within the momentum range relevant for quarkonia our results agree with phenomenological potentials.
We also calculate the propagators for gluons and ghosts in Landau-gauge.\\

\bigskip

\noindent 
PACS-No.s: 12.38.Aw; 12.38.Lg; 12.39.Pn
\end{abstract}
\newpage
\renewcommand{\baselinestretch}{1.7}\normalsize
\section{Introduction}

Quantum Chromodynamics (QCD) is formulated in terms of a functional integral for quarks and gluons.
The observed particles, however, are mesons and baryons.
For a computation of mesonic properties one may imagine a two-step procedure:
First one integrates out the gluons and obtains an effective functional integral for the quarks.
Second one computes the mesonic properties from the effective quark model.
At first sight this seems an extremely difficult task since the result of the first step must be a complicated nonlocal effective action for the quarks.
For example, the first step already contains all the information needed for a determination of the heavy quark potential in the quenched approximation.
Nevertheless, the following four observations indicate that this task may not be insoluble:

{\em{(i)}} Due to asymptotic freedom the momentum dependent multiquark interactions can be computed approximately for momenta sufficiently above typical strong interaction scales,~e.g.~the ``confinement scale'' $\Lambda_{QCD}$.
The full understanding of non-perturbative gluon physics is only needed for the low-momentum behaviour of the quark $n$-point functions which are obtained in the first step.
Furthermore, non-perturbative flow equations based on the scale evolution of the effective average action can extrapolate into a momentum region beyond the perturbative range.
The effective average action $\Gamma_k$ includes in a Euclidean formulation all quantum fluctuations with momenta $p^2$ larger than some infrared cutoff $k^2$.
It is the generating functional for the 1PI $n$-point functions.
A solution for $k \rightarrow 0$ accounts for all quantum effects and therefore contains the relevant information for particle masses and interactions once momenta are analytically continued to Minkowski space.
The dependence of the effective average action on the scale $k$ is described by an exact renormalization group equation
\cite{WegnerWilsonRG,ERGE}.
This allows us to explore the complicated infrared region gradually.

{\em{(ii)}} The functional integration for the quarks in the second step can also be done in small consecutive pieces by the use of exact renormalization group equations.
It has been shown that a formulation in terms of nonperturbative flow equations derived from the effective average action
\cite{ERGE,EAA}
can deal with nonlocal quark interactions
\cite{UliCh}.

Combining this with {\em{(i)}} one expects that a reliable computation of the effective average action for quarks (after integrating out the gluons completely) should be possible at least for momenta and $k$ larger than the vicinity of the confinement scale.

{\em{(iii)}} A first investigation of the flow of the effective average action for quarks has revealed 
\cite{UliCh}
that mesonic bound states typically appear at a ``compositeness scale'' $k_\varphi \sim 600-700$~MeV.
This scale is considerably above the confinement scale and the strong gauge coupling is not yet very large.
For example, the loop expansion of the heavy quark potential still seems to be qualitatively valid for $\sqrt{q^2} \sim 700$~MeV
\cite{HQPot}.
The prospects that at least the physically relevant parts of $\Gamma_{k_\varphi}$ can be computed rather reliably seem therefore quite promising.
This quantity is interesting in its own right since it may be considered as the effective action for Nambu-Jona-Lasinio type models
\cite{NJL}.

{\em{(iv)}} It is possible to rewrite the functional integral such that at the scale $k_\varphi$ the composite mesonic degrees of freedom are included as explicit fields
\cite{UliCh}.
The phenomenon of spontaneous chiral symmetry breakdown occurs as $k$ is lowered beyond $k_\varphi$ (typically for $k \sim 450$~MeV)
\cite{UliCh,DirkCh}.
As a consequence, the quarks acquire a constituent mass $m_q \sim 300 - 350$~MeV and decouple from the flow of the mesonic interactions as $k$ decreases below $m_q$.
The uncertainty in the strong nonperturbative confinement effects expected for the quark $n$-point functions with momenta $q^2 \klgl \Lambda_{QCD}^2$ affects only moderately the meson physics.
Furthermore, the running of many mesonic couplings is governed by a fast attraction towards partial infrared fixed points
\cite{DirkCh}.
Except for a few relevant parameters many properties of the meson physics become therefore rather insensitive to the precise form of $\Gamma_{k_\varphi}$.

This paper aims for a computation of the effective four quark interaction in the first step of such a program.
We will be satisfied with quantitatively reliable estimates for momenta down to $\sqrt{q^2} \sim 300 - 500$ MeV whereas larger uncertainties are tolerable for lower momenta.
More precisely, we want to compute the effective average action for the quarks at a scale $k_\psi$ (typically near the tau lepton or charm quark mass), with the gluon degrees of freedom completely integrated out and the quark fluctuations with momenta larger than $k_\psi$ included.

Our main tool is the effective average action
\cite{ERGE,EAA}.
It is a functional of the fields $\chi$, where the derivatives with respect to $\chi$ are the full or proper vertices.
The average action $\Gamma_k[\chi]$ obtains by adding to the classical action an infrared cutoff piece which is quadratic in the fields
\bea
\Delta_k S = \frac{1}{2} \int \frac{d^d p}{(2 \pi)^d} \chi^\dagger(p) R_k(p) \chi(p)
\label{Null}
\eea
Here $R_k(p^2)$ constitutes a momentum dependent infrared cutoff.
It acts like a mass term $R_k \propto k^2$ for the low momentum modes and is exponentially suppressed for the high momentum modes ($R_k \propto p^2 e^{-p^2/k^2}$ for $p^2 \gg k^2$).
The effective average action $\Gamma_k$ is then computed by the usual steps (introducing sources and performing a Legendre transform) while keeping the infrared cutoff
\cite{ERGE}.
As the infrared cutoff scale $k$ is lowered, quantum fluctuations with larger and larger length scales are effectively included in the functional integration.
The variation of $\Gamma_k$ with $k$ is described by an exact functional differential equation ($t = \ln k/k_0$)
\cite{ERGE}
\bea
\frac{\partial}{\partial t} \Gamma_{k_B,k_F}[\chi] = \frac{1}{2} \STr \left\{ \frac{\partial R_{k_B}^{(B)}}{\partial t} \left( \Gamma^{(2)}_k[\chi] + R_k \right)_{BB}^{-1} \right\} - \Tr \left\{ \frac{\partial R_{k_F}^{(F)}}{\partial t} \left( \Gamma^{(2)}_k[\chi] + R_k \right)_{FF}^{-1} \right\} 
\label{erge}
\eea
In fact, the only $k$-dependence of $\Gamma_k$ arises through the quadratic infrared cutoff (\ref{Null}). 
It is therefore proportional to $\partial R_k / \partial k$ and to the connected two point function $\left( \Gamma_k^{(2)} + R_k \right)^{-1}$. 
One can use different cutoffs $k_B(k)$ and $k_F(k)$ for the bosonic (gluons and ghosts) and fermionic (quarks) parts.
Correspondingly, $\left( \Gamma_k^{(2)}+R_k \right)_{BB}^{-1}$ denotes the projection of the full field dependent propagator on the bosonic subspace and similar for the fermions. 
(For details see reference \cite{ERGE}.)
The traces in eq. (\ref{erge}) contain a momentum integration and a summation over internal and Lorentz indices.
Our exact equation has the form of a renormalization group improved one loop equation.
The only difference from a one loop equation is the appearance of the full inverse propagator $\Gamma_k^{(2)}$,~i.e.~ the matrix of second functional derivatives of $\Gamma_k$ with respect to the bosonic and fermionic fields $\chi$, instead of the classical inverse propagator $S^{(2)}$.
This turns equation (\ref{erge}) into a functional differential equation and renders it exact to all orders in the loop expansion and including all nonperturbative effects.
Furthermore, the momentum integration in equation (\ref{erge}) is both ultraviolet and infrared finite.
Only a narrow range in the loop momenta $p^2 \sim k^2$ contributes effectively.
In consequence, a computation of $\partial_t \Gamma_k$ only needs knowledge of the physics in a momentum range near $k$.

The computation of the fermionic average action at the scale $k_\psi$ can be achieved by starting at some high scale $k_B = k_F = k_0$ within the range of validity of perturbation theory and lowering simultaneously the bosonic and fermionic infrared cutoff ($k_B = k_F = k$) down to $k_\psi$, using equation (\ref{erge}).
One then keeps $k_F = k_\psi$ fixed and removes the infrared cutoff for the bosons ($k_B=k$) by following the evolution towards $k_B \rightarrow 0$.
For this purpose one has $t = \ln k_B / k_0$  and the second term in eq.~(\ref{erge}) does not contribute.
Since the infrared cutoff $k_\psi$ for the quarks is still present all loops containing internal quark lines are suppressed for $k \ll k_\psi$ - similar to the situation for heavy quarks for Euclidean momenta.
The resulting effective action $\Gamma_{0,k_\psi}[\chi]$ is still a functional of both bosonic and fermionic fields.
In order to obtain the effective action involving only quarks $\psi$ one has to solve the bosonic field equations from $\Gamma_{0,k_\psi}$.
The solution expresses the gauge fields as functionals of the fermions.
(The solution for the ghosts corresponds trivially to vanishing ghost fields.)
Inserting this solution into $\Gamma_{0,k_\psi}$ yields the fermionic effective average action $\Gamma_{k_\psi}[\psi]$ which depends only on the quark fields.
We want to compute in this paper the terms in $\Gamma_{k_\psi}[\psi]$ involving two or four powers of $\psi$.
The second step of solving the flow equation for $\Gamma_{k_\psi}[\psi]$ for $k_\psi \rightarrow 0$ is outside the scope of this paper and we refer the reader to 
\cite{UliCh}.

We also relate the effective four quark interaction to the heavy quark potential.
In section 6 we present an approximative estimate of the Fourier transform of the heavy quark potential $V(q^2)$ in the ``quenched approximation'' where no light quarks are included ($N_f=0$).
An optimal renormalization scheme guarantees that $V(q^2)$ almost agrees with the two loop potential in the ${\overline{\mathrm{MS}}}$ scheme for $3 \Lambda_{\overline{\mathrm{MS}}} \leq \sqrt{q^2} \klgl 10 \Lambda_{\overline{\mathrm{MS}}}$.
(Here $\Lambda_{\overline{\mathrm{MS}}} = 315$~MeV is the appropriately adapted two loop confinement scale.)
For lower momenta $\sqrt{q^2} < 2.5 \Lambda_{\overline{\mathrm{MS}}}$ our result for the potential strongly deviates from the two loop potential.
It turns out to be close to phenomenologically motivated potentials as the Richardson potential in the momentum range $\Lambda_{\overline{\mathrm{MS}}} \klgl \sqrt{q^2} \klgl 2.5 \Lambda_{\overline{\mathrm{MS}}}$.
Quarkonia level can therefore be described successfully with our approximate solution of the flow equations.
We also attempt a first estimate of the string tension.
The order of magnitude comes out satisfactory.
The detailed quantitative value stronly depends on the truncation, indicating an insufficiency of the present approximations for very low values of $q^2$.
\begin{figure}[H]
\centering
\epsfig{file=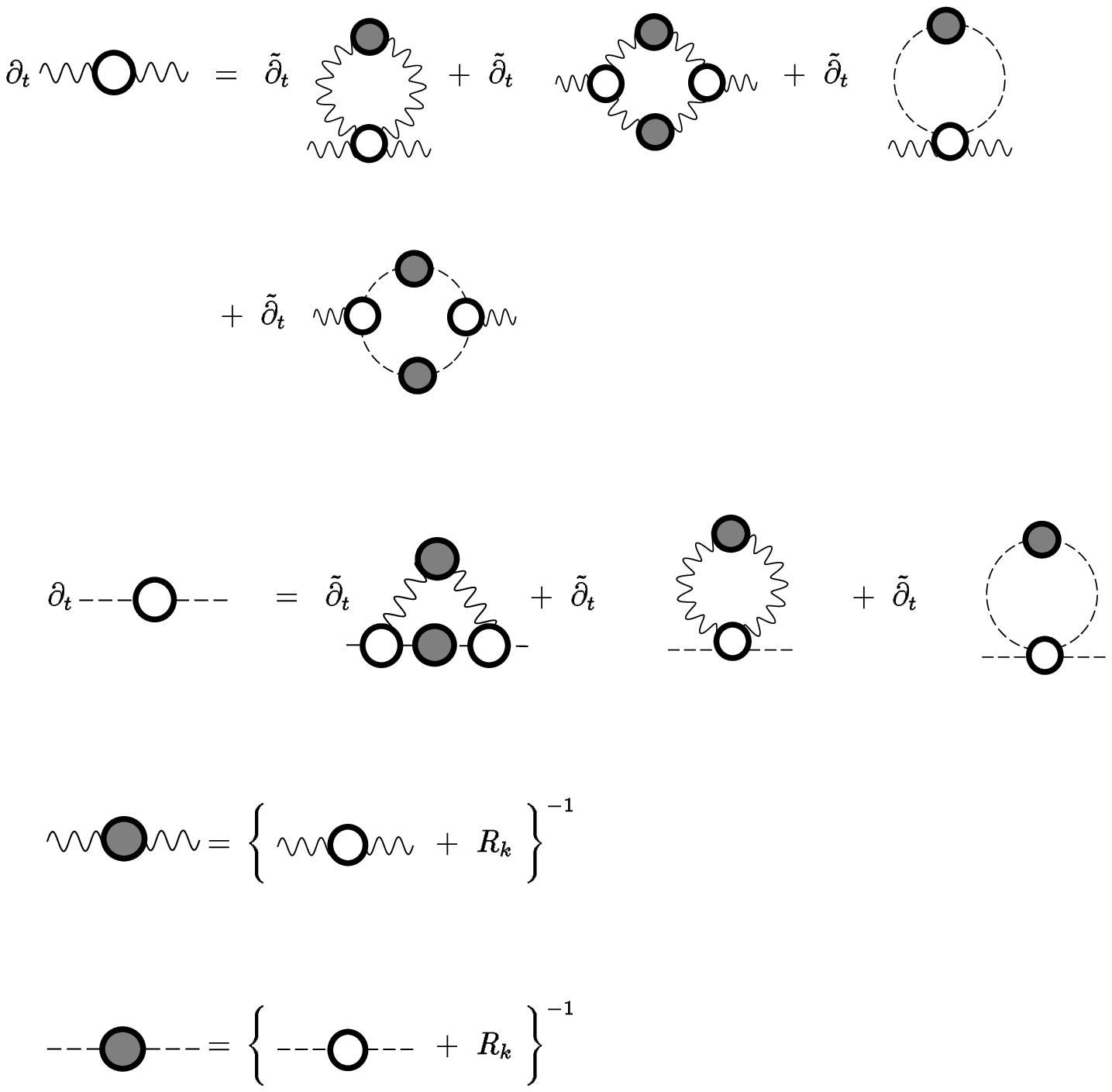,clip=,width=13cm}
\renewcommand{\baselinestretch}{1.2}\normalsize
\caption{ The flow equations for the gluon (wiggly lines) and ghost (broken lines) inverse connected two-point functions $\Gamma_k^{(2)}$. Here $\tilde{\partial}_t = \frac{\partial R_k}{\partial t} \frac{\partial}{\partial R_k}$ denotes a derivative operator that only acts on the explicit $k$-dependence in the cutoff-term. The full dots correspond to the full k-dependent propagators including the cutoff-term whereas the empty dots stand for the full 1PI $n$-point functions. Combinatorical factors are not displayed.}
\renewcommand{\baselinestretch}{1.7}\normalsize
\end{figure}

In the course of computing $\Gamma_{0,k_\psi}[\chi]$ we have to determine the propagators for the gluons and ghosts.
Estimates for these propagators have been given using Schwinger-Dyson equations (SDE)
\cite{SDERev}.
In a certain sense our non-perturbative flow equations can be interpreted as a differential form of the SDE.
In contrast to the SDE only a narrow momentum window $p^2 \sim k^2$ contributes effectively to the momentum integration in the flow equation (\ref{erge}) and truncations should therefore be more reliable.
For a possible comparison with other approaches we also present in this paper an estimate of the gluon and ghost propagators in Landau gauge and the quenched approximation.
\begin{figure}[H]
\begin{center}
\epsfig{file=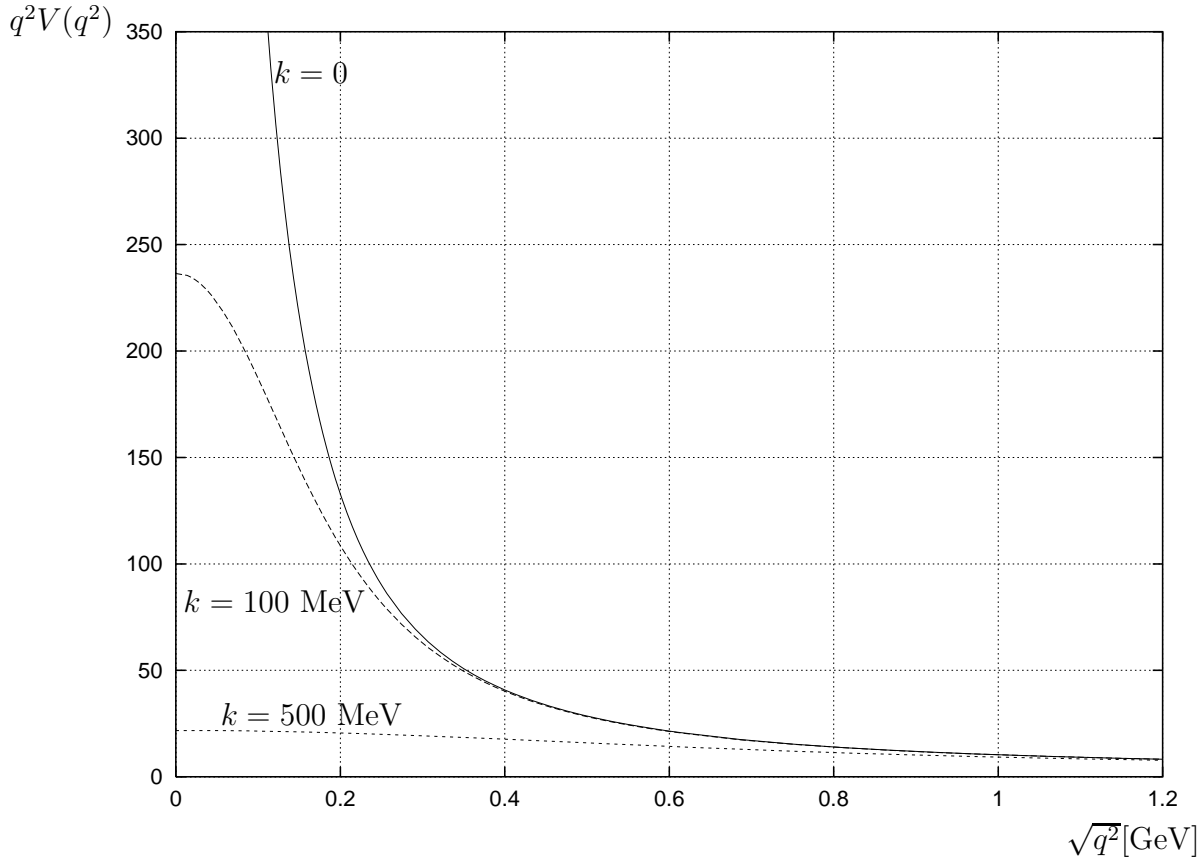,width=15cm}
\put (-450,295) {$q^2 V(q^2)$}
\put (-50,-15) {$\sqrt{q^2} [\mathrm{GeV}]$}
\put (-350,275) {$k=0$}
\put (-384,75) {$k=100$~MeV}
\put (-370,32) {$k=500$~MeV}\\
\end{center}
\renewcommand{\baselinestretch}{1.2}\normalsize
\caption{Momentum dependence of $q^2 V(q^2)$ for different values of the cutoff $k$.
The dotted line corresponds to $k = 500$~MeV, the dashed line is for $k = 100$~MeV, and the solid line displays the final result for $k=0$.}
\renewcommand{\baselinestretch}{1.7}\normalsize
\end{figure}

In practice, the most important ingredient of our calculation is the numerical solution of non-perturbative flow equations for the gluon and ghost propagators.
These equations obtain from the exact equation (\ref{erge}) by a truncation which is explained in detail in the next section.
Graphically, the exact system of flow equations for the gluon and ghost propagator (for fixed $k_F = k_\psi$) is represented in figure 1.
The truncation corresponds to an approximation for the vertices and the precise form of the flow equation can be found in section 3 (equations (\ref{dtGA}) and (\ref{dtGgh})).
As the infrared cutoff $k \equiv k_B$ is lowered, more and more of the fluctuations are included.
This results in a transition from an almost classical form of the potential $V \propto 1/q^2$ for $k = 500$~MeV to a non-perturbative form for $k \rightarrow 0$, as can be seen in figure 2.

In order to connect our results for small $q^2$ with the perturbative results for large $q^2$ it is important to choose the initial values for the numerical solution of the flow equation carefully.
Otherwise, the overall (non-perturbative) mass scale appearing in the potential for the range $300~{\mathrm{MeV}} < \sqrt{q^2} < 700~{\mathrm{MeV}}$ would remain a free parameter.
We discuss in detail the choice of initial conditions in sections 4 and 5.
This allows us to establish an explicit relationship of the relevant mass scale with the perturbative mass scale appearing in the $\overline{\mathrm{MS}}$ scheme.

We should point out that the present paper is closely related to earlier work in the context of exact renormalization group equations on QCD propagators and the heavy quark potential
\cite{ChQCD,UliManfredAxel},
even though the present perspective is somewhat different.

\section{Effective four-quark interaction} 

In this work we compute the effective four quark interaction in $\Gamma_{0,k_\psi}$.
We always assume here $k_F = k_\psi$ held fixed and identify $k = k_B$.
In order to approximately solve the exact flow-equation (\ref{erge}) we have to resort to a truncation of the most general form of the effective average action.
We work in the general setting of flow equations for gauge theories
\cite{MartinCh,Becchi,Italiener,Uli}
with Landau gauge fixing.
This is motivated by the fact that a vanishing gauge fixing parameter $\alpha=0$ is a fixed point of the flow equations
\cite{ChQCD,ulietal}
and will turn out to be crucial later.
We work with vanishing ``background fields'' where the effective average action is not a gauge invariant functional.
As far as the Slavnov-Taylor identities are concerned one notes
\cite{Uli}
that \Gk~contains counterterms, $\Gamma_{ct}$, which are induced by the breaking of BRS-invariance due to the infrared cutoff $R_k$.
For $k_B, k_F \rightarrow 0$ the BRS invariance is restored and the counterterms vanish.
We may implicitly define $\Gamma_{ct}$ by requiring that $\Gamma_k-\Gamma_{ct}$ obeys the standard Slavnov-Taylor identities.
An exact solution of these identities which is sufficiently general for our purpose can be written in the form \cite{ChQCD} ($\alpha \rightarrow 0$)
\bea
\Gamma_k\!\!&=&\!\!\Gamma_{inv}[\hat{A},\psi] - \int d^4x\,\,\hat{\bar{\xi}}^z \partial^\mu \left( D_\mu(\hat{A}) \xi \right)^z+ \frac{1}{2\alpha} \int d^4x \left( \partial^\mu A^z_\mu \right) \left( \partial^\nu A^z_\nu \right) + \Gamma_{ct}
\label{general_solution_to_BRS}
\eea
Here $\Gamma_{inv}$ is at this point an arbitrary gauge invariant functional of the rescaled gauge field $\hat{A}$ and the quark fields $\psi$, with covariant derivative $D_\mu(\hat{A}) \psi = \partial_\mu \psi - i \bar{g} T_z \hat{A}^z_\mu \psi$.
The fields $\hat{A}$ and $\hat{\bar{\xi}}$ are related to the gauge field $A$ and the antighost $\bar{\xi}$ by a momentum dependent wave-function renormalization
\be
\hat{A} = H(-\partial^2) A \qquad , \qquad \hat{\bar{\xi}} = H^{-1}(-\partial^2) \bar{\xi}
\label{relation_of_fields}
\ee
Comparison with (\ref{general_solution_to_BRS}) relates $H$ to the inverse ghost propagator $G_{gh}$ by
\be
G_{gh}(q^2) = q^2 H^{-1}(q^2)
\label{relation_Ggh_and_H}
\ee
The function $H(q^2)$ will depend on $k$ and the scaling of $\hat{A}$ is such that the covariant derivative always involves a fixed gauge coupling $\bar{g}$.
This brings the present formulation close to the background field formulation \cite{MartinCh}, even though we will work here with a vanishing background field.
From the ghost sector in our truncation we only need information about the two-point function $G_{gh}(q^2)$, since the $\hat{A}\hat{\bar{\xi}}\xi$-vertex is fixed by the constant $\bar{g}$ in Landau-gauge.

In this setting the effective quark interactions can be extracted from the gauge invariant kernel $\Gamma_{inv}[\hat{A},\psi]$. 
For the general discussion we will consider the truncation{\footnote{We work in Euclidean space and $\bar{\gamma}$ is the Euclidean analogue of $\gamma_5$ (for conventions see 
\cite{HD-THEP-95-2}).}}
\bea
\Gamma_{inv}\!\!&=&\!\!\int d^4 x \Biggl \lbrace \frac{1}{4}\hat{F}^z_{\mu\nu} \left( Z_F(-D^2) \hat{F}^{\mu\nu}\right)^z +\frac{1}{2}\left(D^\rho\hat{F}_{\mu \rho}\right)^z Y_F(-D^2) \left( D_\nu \hat{F}^{\mu \nu}\right)^z+ \nonumber \\
&& \hspace*{2cm}+ \sum_a \bar{\psi}_a \left( i\gamma^\mu D_\mu Z_\psi^a(-D^2) + \bar{m}_\psi^a \bar{\gamma} + \gamma^\mu\left[Y_\psi^a(-D^2)\left(D^\nu F_{\mu\nu}\right)\right]^zT_z \right)\psi_a - \nonumber \\
&& \hspace*{2cm}- \sum_{a,b}\frac{1}{2} \left( \bar{\psi}_a \gamma^\mu T_z \psi_a \right) E^{ab}(-D^2) \left( \delta_{\mu\nu} - \frac{D_\mu D_\nu}{D^2} \right)\left( \bar{\psi}_b \gamma^\nu T_z \psi_b \right) \Biggr \rbrace
\label{first_truncation}
\eea
which will be further restricted below for our practical calculation.
Here $Z_F$, $Y_F$, $Z_\psi^a$, $Y_\psi^a$, and $E^{ab}$ are arbitrary functions of the covariant Laplacian (involving $\hat{A}$) in the appropriate representation and $\hat{F}^z_{\mu\nu} = \partial_\mu \hat{A}^z_\nu - \partial_\nu \hat{A}^z_\mu + \bar{g}f^{zxy} \hat{A}^x_\mu \hat{A}^y_\nu$.
The index $a$ labels the different quark-flavors.
We denote the contribution from $\Gamma_{inv}$ to the transversal two-point function for $A$ (not $\hat{A}$) by $\overline{G}_A(q^2)$,
\bea
\overline{G}_A(q^2) = \left( q^2 Z_F(q^2)+q^4 Y_F(q^2) \right) H^2(q^2) = \left( Z_F(q^2)+q^2 Y_F(q^2) \right) q^6 G_{gh}^{-2}(q^2) 
\label{relation_ZF_and_Gs}
\eea
The last term in equation (\ref{first_truncation}) $\propto E$ accounts for part of the 1PI four quark Green-function.

In order to extract the information on effective quark-interactions we have to solve the field-equation for the gluon-field as a functional of the quark field and insert the solution into the effective action. 
Restricting ourselves to the four-quark interaction, the solution of the field-equation for $A$ derived from (\ref{first_truncation}) is needed only in lowest order in an expansion in $(\bar{\psi} \psi)$ 
\cite{ChQCD}.
Upon inserting the solution into $\Gamma_{inv}[A,\psi]$ we find $\Gamma_{eff}[\psi]$ as{\footnote{
The additional $\bar{\psi}_a\psi_a\hat{A}$-vertex connected to a non-constant $Z_\psi^a(-D^2)$ contributes to the field-equation for $A$, but leads to a different index structure for the four-quark interaction.
This also holds for gauge invariants of the type $\bar{\psi}_a\gamma^\mu\gamma^\nu\hat{F}^z_{\mu\nu}T_z\psi_a$ which are not contained in our truncation.
Nevertheless, there are in general also neglected invariants contributing to the index structure $(\bar{\psi}_a\gamma^\mu \psi_a)^2$.}
\bea
\Gamma_{eff}[\psi] &=& \int d^4x \Biggl \lbrace \sum_a \bar{\psi}_a \left( i \gamma^\mu \partial_\mu Z^a_\psi(-\partial^2) + \bar{m}_\psi^a \bar{\gamma} \right) \psi_a - \frac{1}{2} \sum_{a,b} \left(\bar{\psi}_a \gamma^\mu T_z Z^a_\psi(-\partial^2) \psi_a\right) \times \nonumber \\
&& \times \left[ \frac{(\bar{g} - \partial^2 Y_\psi(-\partial^2))^2}{-\partial^2 (Z_F(-\partial^2) - \partial^2 Y_F(-\partial^2))} + E(-\partial^2) \right]^{ab} \left( \delta_{\mu\nu} - \frac{\partial_\mu \partial_\nu}{\partial^2} \right) \times \nonumber\\ 
&& \qquad \quad \times \left(\bar{\psi}_b \gamma^\nu T^z Z^b_\psi(-\partial^2) \psi_b\right) + ... \Biggr \rbrace
\label{quark_action}
\eea
where we have neglected terms with different index-structures for the four-quark interaction and higher powers of $(\bar{\psi} \psi)$.
We will concentrate on the quantity
\bea
\hat{V}(q^2)&=& \frac{4}{3} \left[\frac{ \left(\bar{g} + q^2 Y_\psi(q^2) \right)^2}{\left(Z_F(q^2)+q^2 Y_F(q^2)\right) q^2} +E(q^2) \right] \nonumber\\
&\simeq& \frac{4}{3} \frac{\bar{g}^2 q^4}{\overline{G}_A(q^2) G^2_{gh}(q^2)} + \frac{\tilde{\Delta}}{q^2} \equiv \tilde{V}(q^2) + \frac{\tilde{\Delta}}{q^2}
\label{V}
\eea
where the normalization is such that $\hat{V}(q^2)$ coincides with the Fourier transform of the heavy quark potential\footnote{Here $V(q^2)$ corresponds to the quenched approximation without light quarks ($N_f=0$).} $V(q^2)$ in the one loop approximation up to terms $\propto g^4/q^2$. (In equation (\ref{V}) we have suppressed the flavor indices of $Y_\psi$ and $E$).
We want to compute $\hat{V}(q^2)$ in the limit $k\rightarrow 0$ such that all gluon-fluctuations are integrated out.
We will also henceforth approximate\footnote{The solution of the modified Slavnov-Taylor Identities in ref.
\cite{UliManfredAxel}
is not general enough to admit nonvanishing $E$ or $Y_\psi$.
Note that in general gauges the expression for $\tilde{V}$ does not reproduce the correct running of the gauge coupling even to one-loop. 
This is due to contributions that are absent in Landau gauge.} the contributions from $Y_\psi$ and $E$ in the Landau gauge by a $k$-independent constant $\tilde{\Delta} \propto \bar{g}^4$.
This is suggested by the observation that the running of $Y_\psi$ and $E$ involves loops with internal fermion lines and is suppressed for a fixed fermionic cutoff $k_F$.
In summary, our calculation amounts to the determination of the scale dependent gluon and ghost propagators $G_A(q^2)$ and $G_{gh}(q^2)$. 

For $k>0$ we also have to specify our truncation for the counterterms contained in $\Gamma_{ct}$. 
In principle, this piece is completely fixed by the additional terms appearing in the Slavnov-Taylor identities for $k>0$ \cite{Uli}.
We only include explicitly{\footnote{The contribution from $\Gamma_{ct}$ to the gluon two-point function $G_A(q^2)$ also includes $q^2$-dependent terms.
The flow-equation for $\overline{G}_A(q^2)$ presented below actually describes the evolution of the full gluon two-point function derived from $\Gamma_{inv}+\Gamma_{ct}$, with the constant for zero momentum (the massterm $\overline{m}_A^2$) subtracted such that $\overline{G}_A(0)\equiv 0$. 
The final result for $\overline{G}_A(q^2)$ for $k=0$ nevertheless describes only $\Gamma_{inv}$, since the contribution from $\Gamma_{ct}$ vanishes.
For the approximation of the vertices used in this work there is no need to separate the contributions from $\Gamma_{inv}$ and $\Gamma_{ct}$.
Thus the $q^2$-dependent part in $\Gamma_{ct}$ is implicitly included.
Only the mass term needs to be determined in addition to $\overline{G}_A(q^2)$, since by definition $\overline{G}_A(q^2) = G_A(q^2)-G_A(0) = G_A(q^2)-\overline{m}_A^2$.
Similar remarks apply to $G_{gh}$ where we note that $G_{gh}(0)$ remains identically zero (cf.~also eq.~(\ref{dtGgh})) such that $\overline{m}_{gh}^2 = 0$ for all $k$.}}
a gluon mass term, given by
\bea
\overline{m}_A^2\!\!&=&\!\!\tilde{Z}_A\frac{3 g_k^2}{16 \pi^2} \int_0^\infty dx x \Biggl \lbrace
-\frac{15}{4}\frac{\tilde{Z}_A R_k^{(A)}(x)}{\left( G_A(x)+R_k^{(A)}(x) \right)^2} + 3 x \frac{\tilde{Z}_A R_k^{(A)}(x) \frac{d}{dx} \left( G_A(x)+R_k^{(A)}(x) \right)}{\left( G_A(x)+R_k^{(A)}(x) \right)^3} + \nonumber \\
&& \qquad \qquad +\frac{\tilde{Z}_{gh} R_k^{(gh)}(x)}{\left( G_{gh}(x)+R_k^{(gh)}(x) \right)^2} -\frac{1}{2} x \frac{\tilde{Z}_{gh} R_k^{(gh)}(x) \frac{d}{dx} \left( G_{gh}(x)+R_k^{(gh)}(x) \right)}{\left( G_{gh}(x)+R_k^{(gh)}(x) \right)^3} + \nonumber \\
&&  \qquad \qquad +\frac{\tilde{Z}_{gh} R_k^{(A)}(x)}{\left( G_A(x)+R_k^{(A)}(x) \right) \left( G_{gh}(x) + R_k^{(gh)}(x)\right)}- \nonumber \\
&& \qquad \qquad - \frac{1}{2} x \frac{\tilde{Z}_{gh} R_k^{(A)}(x) \frac{d}{dx} \left( G_{gh}(x)+R_k^{(gh)}(x) \right)}{\left( G_A(x)+R_k^{(A)}(x) \right)\left( G_{gh}(x) + R_k^{(gh)}(x)\right)^2} \Biggr \rbrace
\label{gluon_mass}
\eea
Here the wave function renormalization constants are given by
\bea
\tilde{Z}_A(k)=\left. \frac{\partial \overline{G}_A(q^2)}{\partial q^2}\right|_{q^2=0} \qquad ; \qquad \tilde{Z}_{gh}(k)=\left. \frac{\partial G_{gh}(q^2)}{\partial q^2}\right|_{q^2=0}  
\label{WFRs}
\eea
In the flow of general models for vector fields this mass term would be a relevant parameter, whereas in gauge theories its value is dictated by symmetry.
Computing the mass term from the flow of truncated evolution equations could easily lead to large deviations from the ``gauge theory hypersurface'' in parameter space.
It is therefore preferable to determine the mass term directly from the appropriate symmetry relations as in equation (\ref{gluon_mass}).
As the infrared cutoff is removed the mass term vanishes for $k \rightarrow 0$.
The mass term depends on the specific form of the infrared cutoff-function $R_k$. 
A natural choice for $R_k(p^2)$ is
\bea
R_k(p^2) = \tilde{Z} p^2 \frac{e^{-p^2/k^2}}{1-e^{-p^2/k^2}}
\label{R_k}
\eea
with different $k$-dependent wave function renormalization functions $\tilde{Z}$ for gluons and ghosts.
In the region of approximately perturbative behaviour for $k > 3 \Lambda_{\overline{\mathrm{MS}}}$ (with $\Lambda_{\overline{\mathrm{MS}}}$ the two loop confinement scale in the $\overline{\mathrm{MS}}$ scheme) there is also approximate scaling in the renormalized quantities and we choose $k$-dependent $\tilde{Z}$ in $R_k$ by identifying the constants in the infrared cutoff with the running wave function renormalizations (\ref{WFRs}) for the gauge fields and ghosts, respectively.
For $k < 3 \Lambda_{\overline{\mathrm{MS}}}$ scale invariance is strongly violated anyhow and we do not change $\tilde{Z}$ with $k$ anymore, i.e.~we take $\tilde{Z} = \tilde{Z}(3 \Lambda_{\overline{\mathrm{MS}}})$ in $R_k$.
For low momenta $p^2 \ll k^2$ the infared cutoff $R_k$ acts like a mass term $\propto \tilde{Z} k^2$. 
Typically, the size of the mass term contained in $\Gamma_{ct}$ is smaller than $\tilde{Z} k^2$.
In a certain sense, the mass term $\propto k^2$ actys similar to a modification of the infrared cutoff $R_k$.
Including the contributions from $\Gamma_{ct}$ the $k$-dependent inverse transverse gluon-propagator $G_A$ obeys $G_A(q^2) = \overline{G}_A(q^2) + \overline{m}_A^2$.

\section{Flow equation}

Our method amounts to a solution of flow-equations for the $k$-dependence of the gluon and ghost two-point functions $\overline{G}_A(q^2)$ and $G_{gh}(q^2)$, starting at some high momentum scale $k=k_0$ with perturbative ``initial conditions'' and extrapolating to $k \rightarrow 0$.
The exact flow-equations for the two-point functions obtain by taking the second functional derivative of equation (\ref{erge}) and therefore involve the exact momentum dependent three- and four-point functions between gluons, quarks and ghosts. 
Those are not known and we have to proceed to some approximation.
(In principle, the general flow equation (\ref{erge}) also yields exact flow equations for the three and four point functions. 
Since the latter involve again higher $n$-point functions a solution of such equations would postpone the necessity of approximations to the higher $n$-point functions.)
One possible approach would be the extraction of the three and four-point functions from the truncation (\ref{general_solution_to_BRS}), (\ref{first_truncation}), expressing them in terms of the functions\footnote{See 
\cite{UliManfredAxel}
for the limit where $Y_F$ is neglected.} $Z_F$, $Y_F$ and $H$.
There is, however, no reason to believe that such a procedure is particularly accurate - it is easy to construct additional invariants contributing to the three and four-point functions as, for example, $\left( \hat{F}_{\mu\nu} \hat{F}^{\mu\nu} \right)^2$.
Instead of such a truncation we will base our approximation on the observation that only a relatively narrow range in the integration momentum $p^2$ contributes effectively to the integrals on the right hand sides of the flow equations for $\partial_t G_A(q^2)$ and $\partial_t G_{gh}(q^2)$.
For a given propagator momentum $q^2$ there should exist some effective gauge coupling $g_k$ such that a classical approximation of the renormalized vertices gives the same result as the full momentum dependent three and four-point functions.
(This statement is close in spirit to the central value theorem for ordinary integration.)
Of course, this effective coupling may be different for different values of $q^2$ and between $\partial_t G_A(q^2)$ and $\partial_t G_{gh}(q^2)$.
Our approximation consists now in taking $g_k$ independent of the momentum of the propagators and using the same value for $\partial_t G_A(q^2)$ and $\partial_t G_{gh}(q^2)$.
In practice this permits the use of the classical momentum dependence for the three and four-point functions on the right hand side of the flow equations.
We should point out, however, that this does by no means imply the assumption that the vertices are close to the classical ones.
We only suppose that a certain momentum weighted average can be represented by a similar average over classical vertices with appropriate $g_k$.
It is a drawback of this approach that in practice $g_k$ is only defined implicitly\footnote{It is not difficult, however, to write down a formal expression in terms of momentum weighted integrals over exact three and four point functions.}, without an explicit representation in terms of $n$-point vertices.
This results in an uncertainty about what should be the scale dependence of $g_k$ and we will specify our assumptions below.
In fact, we believe that the main uncertainty concerns this scale dependence.
The approximation of an effective $q^2$-independence of $g_k$ probably plays only a minor role, at least for momenta above the confinement scale.

In summary, we insert on the right hand side of the flow equations a classical momentum dependence for the three point functions and a constant four point function in terms of renormalized gauge fields $A_R=\tilde{Z}_A^{1/2} A$ and ghosts $\bar{\xi}_R = \tilde{Z}_{gh}^{1/2}\bar{\xi}$, $\xi_R = \tilde{Z}_{gh}^{1/2} \xi$.
The strength of the renormalized vertices is then determined by a common renormalized gauge coupling $g_k$ appearing in the covariant derivative $\partial_\mu - i g_k T^z A_{R\mu}^z$. 
Our approximation can be compared with the so-called Mandelstam-approximation in the SDE approach 
\cite{Mandelstam} with the important difference that in contrast to the SDE only a narrow momentum range contributes in the loop integrals involved in the flow equations.
An effective ``momentum averaging of vertices'' should therefore work much better in our case. 

With our approximations the flow-equations for the two-point functions simplify considerably and one finds for Landau gauge-fixing $\alpha=0$
\cite{ChQCD}
\bea
\frac{\partial}{\partial t}\overline{G}_A(q^2)\!\!&=&\!\!
3 g_k^2 \int\frac{d^4p}
{(2\pi)^4}\tilde\partial_t\Biggl\lbrace
2\tilde{Z}^3_A\left(\overline{G}_A(p)\!+\!\overline{m}_A^2\!+\!R_k(p)\right)^{-2}\left({p}^2-\frac{(qp)^2}{q^2}\right)- \nonumber \\
&& \qquad \qquad \qquad 
-\frac{1}{3}\tilde{Z}^2_{gh}\tilde{Z}_A\left(G_{gh}(p)\!+\!R_k(p)\right)^{-2}\left({p}^2-\frac{(qp)^2}{q^2}\right)-
\nonumber\\
&&-\frac{1}{6}\tilde{Z}_A^3\left(\overline{G}_A(p)\!+\!\overline{m}_A^2\!+\!R_k(p)\right)^{-1}
\left(\overline{G}_A(q\!+\!p)\!+\!\overline{m}_A^2\!+\!R_k(q\!+\!p)\right)^{-1}\times\nonumber\\
&&\times\left[13 q^2 - 14 (qp) + 12 {p}^2 - 12 \frac{(qp)^2}{q^2}
\right.-\left. 22 \frac{(qp)^2}{{p}^2} - 4 \frac{(qp)^3}{q^2{p}^2} + \frac{q^2}{{p}^2}\frac{q^2{p}^2-(qp)^2}{(q+p)^2}\right]+
\nonumber\\
&&+\frac{1}{3}\tilde{Z}_{gh}^2\tilde{Z}_A\left(G_{gh}(p)\!+\!R_k(p)\right)^{-1}\left(G_{gh}(q\!+\!p)\!+\!R_k(q\!+\!p)\right)^{-1}
\left[{p}^2-\frac{(qp)^2}{q^2}\right] \Biggr\rbrace
\label{dtGA}
\eea
\bea
\frac{\partial}{\partial t}G_{gh}(q^2)\!\!&=&\!\!- 3 g_k^2 \tilde{Z}_A \tilde{Z}^2 _{gh}\int\!\frac{d^4p}{(2\pi)^4}\tilde\partial_t 
\Biggl\lbrace \left(\overline{G}_A(p)\!+\!\overline{m}_A^2\!+\!R_k(p)\right)^{-1}\! \left(G_{gh}(q\!+\!p)\!+\!R_k(q\!+\!p)\right)^{-1}\!\times\nonumber\\ 
&& \qquad \qquad \times \left[q^2-\frac{(qp)^2}{p^2}\right]\Biggr\rbrace 
\label{dtGgh}
\eea
Here the symbol $\tilde{\partial}_t$ denotes a $t$-derivative which acts only on the infrared cutoff $R_k$, i.e.~$\tilde{\partial}_t= (\partial R_k/\partial t)(\partial/\partial R_k)$.
The exponential decay of $\partial_t R_k$ (cf. equation (\ref{R_k})) for large $p^2 / k^2$ implies a fast ultraviolet convergence of the momentum integrals in (\ref{dtGA}) and (\ref{dtGgh}).

It remains to specify the scale-dependence of the coupling $g_k$.
Even though defined only implicitly, $g_k$ is a possible choice of a renormalized gauge coupling (in an appropriate implicitly defined scheme).
For small $g_k$ its running must therefore be governed by the universal two-loop beta-function and we use for sufficiently large $k$
\bea
\frac{\partial}{\partial t} g^2_k = \beta_{g^2}^{(2)}(g_k^2) = - 22 \frac{g_k^4}{16 \pi^2} - 204 \frac{g_k^6}{(16 \pi^2)^2} \qquad \quad {\mbox{for}}~k>k_c
\label{beta_g}
\eea
(No quark loops are included here since the fermionic cutoff $k_F$ remains fixed.
The running is then given by QCD with only heavy quarks or an effective number of light flavors $N_f=0$.)
For larger $g_k$ we have in the present approximations no knowledge about the evolution of $g_k$.
Also our approximation of a classical momentum dependence for the vertices becomes less accurate.
We find it suitable to display the effect of these uncertainties by using a set of different $\beta$-functions for large $g_k$.
We replace (\ref{beta_g}) for $k<k_c$ by 
\bea
\frac{\partial}{\partial t} g_k^2 = c (g_k^2-g_\ast^2) \quad ; \quad  c=\frac{\beta_{g^2}^{(2)}(g_c^2)}{g_c^2-g_\ast^2} \quad , \quad g_c^2 = g_k^2(k_c) \qquad {\mbox{for}}~k<k_c
\label{beta_g_small}
\eea
This has the effect that the increase of $g_k$ is smoothly stopped for small $k$.
For $k \rightarrow 0$  $g_k$ approaches a fixed value $g_\star$.
The scale where the running of $g_k$ starts deviating from the two loop running is fixed by the condition $g_c^2 = 0.95 g_\star^2$.
Equation (\ref{beta_g_small}) corresponds then to a one parameter family of $\beta$-functions.
We will parameterize the uncertainty of our approximation by 
\bea
\alpha_\star = \frac{g_\star^2}{4 \pi}
\label{IRalpha}
\eea
with typical values of $\alpha_\star$ of order one.
It will become apparent that our lack of knowledge of the exact vertices plays an important role for the shape of $V(q^2)$ only at small $q^2$.
For $\sqrt{q^2} \grgl 600$ MeV the dependence of $V(q^2)$ on $\alpha_\star$ turns out to be weak.

After performing the $p$ integrations, the flow equations (\ref{dtGA}) and (\ref{dtGgh}) become a coupled system of nonlinear partial differential equations for two functions $\overline{G}_A$, $G_{gh}$, depending on two variables $q^2$ and $t$.
They are supplemented by equations (\ref{beta_g}), (\ref{beta_g_small}) for the running of $g_k$.
We have solved these equations numerically, starting at $k_0$ from perturbative initial conditions for $\overline{G}_A(q^2)$ and $G_{gh}(q^2)$.

\section{Initial Values}

The ``initial values'' for the propagators at the scale $k_0$ are determined by the solution of the flow equations for $k \geq k_0$, starting from very large $k$ with the ``classical action'' including appropriate counterterms $\Gamma_{ct}$.
Since for $k > k_0$ perturbation theory is a valid expansion we can treat the flow of ${\overline{G}}_A$ and $G_{gh}$ for $k>k_0$ perturbatively.
The flow-equations in the one-loop approximation are obtained by setting on the right-hand sides of (\ref{dtGA}) and (\ref{dtGgh}) $\overline{G}_A(q^2) = G_{gh}(q^2) = q^2$, $\tilde{Z}_i = 1$, $\overline{m}_A^2 = 0$ and $g_k^2 = \bar{g}^2$.
In this case the momentum-integrations can be performed explicitly.
We find
\bea
\frac{\partial \overline{G}_A(q^2)}{\partial t}\!&=&\!\frac{\bar{g}^2 k^2}{16 \pi^2} \Biggl \lbrace \frac{27}{4}\!+\!\frac{44}{x}\!-\!\frac{15}{x^2}\!+\!\frac{7x^2}{2}\!\left( {\mbox{Ei}}_1\!(x)\!-\!{\mbox{Ei}}_1\!(x/2) \right)\!-\!x^3\!\left( {\mbox{Ei}}_1\!(x)\!-\!\frac{{\mbox{Ei}}_1\!(x/2)}{2} \right)\!+ \nonumber \\
&& \quad + e^{-x}\left(\!x^2\!-\!\frac{9x}{2}\!+\!\frac{11}{2}\!+\!\frac{23}{x}\!+\!\frac{9}{x^2} \right)\!+ e^{-\frac{x}{2}} \left(\!-x^2\!+\!9 x\!-\!22\!-\!\frac{55}{x}\!+\!\frac{6}{x^2} \right) \Biggr \rbrace\!+\!{\mathcal{O}}(\bar{g}^4)
\label{dtGA_1loop}
\eea
and
\bea
\frac{\partial G_{gh}(q^2)}{\partial t}&=&\frac{\bar{g}^2 k^2}{16 \pi^2} \Biggl \lbrace \frac{9}{2} + \frac{6}{x} + \frac{3}{2} x^2 \left( {\mbox{Ei}}_1\!(x) - {\mbox{Ei}}_1\!(x/2) \right) + \exp(-x)\left( - \frac{3x}{2}+ \frac{3}{2} + \frac{6}{x} \right) + \nonumber \\
&& \quad + \exp(-x/2) \left( 3 x - 6 - \frac{12}{x} \right) \Biggr \rbrace + {\mathcal{O}}(\bar{g}^4)
\label{dtGgh_1loop}
\eea
where $x = q^2/k^2$ and 
\bea
{\mbox{Ei}}_1\!(x) = \int_1^\infty dt \frac{e^{-xt}}{t} \nonumber
\label{Ei}
\eea
Note that the $q^2$-derivatives of the above expressions reproduce for $q^2 \rightarrow 0$ the perturbative anomalous dimensions as they should.

The coupling $\bar{g}^2$ does not depend on $k$ and we can integrate equations (\ref{dtGA_1loop}) and (\ref{dtGgh_1loop}).
This yields at $k=k_0$ the one-loop expressions\footnote{The integration of equations (\ref{dtGA_1loop}) and (\ref{dtGgh_1loop}) should be performed between $k_0$ and some ultraviolet scale $\Lambda_0 \gg k_0$.
For $\Lambda_0 / k_0 \rightarrow \infty$ the contribution from the upper boundary $\Lambda_0$ reduces to a simple term $\propto {\mbox{Ei}}_1\!(\frac{q^2}{2 \Lambda_0^2}) \rightarrow - \left( \ln x + \ln \frac{k_0^2}{\Lambda_0^2} \right)$.
This yields the logarithms in equations (\ref{GA_1loop}) and (\ref{Ggh_1loop}) and a $k_0$-dependent piece in the constants, $z_A = {\mathrm{const.}} - \frac{13 \bar{g}^2}{16 \pi^2} \ln \frac{\Lambda_0}{k_0}$ and similar for $z_{gh}$.} ($x=q^2/k_0^2$)
\bea
\frac{\overline{G}_{A,k_0}(q^2)}{q^2}&=&z_A + \frac{3 \bar{g}^2}{16 \pi^2} \Biggl \lbrace \frac{13}{6} \left( \ln x + {\mbox{Ei}}_1\!(x/2) \right) + \frac{9}{8x}+\frac{11}{3x^2}-\frac{5}{6x^3} - \frac{7x}{12} \left( {\mbox{Ei}}_1\!(x) - {\mbox{Ei}}_1\!(x/2) \right) + \nonumber\\
&& \qquad \quad + \frac{x^2}{12} \left( {\mbox{Ei}}_1\!(x) - \frac{{\mbox{Ei}}_1\!(x/2)}{2} \right) + e^{-x} \left( -\frac{x}{12} + \frac{2}{3} - \frac{3}{4x} + \frac{5}{3x^2} + \frac{1}{2x^3} \right) + \nonumber \\
&& \qquad \quad + e^{-x/2} \left( \frac{x}{12} - \frac{4}{3} - \frac{4}{3x} - \frac{14}{3x^2} + \frac{1}{3x^3} \right) \Biggr \rbrace
\label{GA_1loop}
\eea
and
\bea
\frac{G_{gh,k_0}(q^2)}{q^2}&=&z_{gh} + \frac{3 \bar{g}^2}{16 \pi^2} \Biggl \lbrace \frac{3}{4} \left( \ln x + {\mbox{Ei}}_1\!(x/2) \right) + \frac{3}{4x}+\frac{1}{2x^2} - \frac{x}{4} \left( {\mbox{Ei}}_1\!(x) - {\mbox{Ei}}_1\!(x/2) \right) + \nonumber \\
&& \qquad \quad +e^{-x} \left( \frac{1}{4} - \frac{1}{4x} + \frac{1}{2x^2} \right) - e^{-x/2} \left( \frac{1}{2} +\frac{1}{2x} +\frac{1}{x^2} \right) \Biggr \rbrace
\label{Ggh_1loop}
\eea
The integration constants $z_A$ and $z_{gh}$ are related to the choice of a normalization scheme (see next section). 
For the gluon-mass counterterm we find to one loop 
\cite{Uli,ChQCD}
\bea
\bar{m}_A^2 = - \frac{9}{8} \frac{\bar{g}^2}{16 \pi^2} k^2 
\label{oneloopmass}
\eea
We have also used equations (\ref{dtGA_1loop}) - (\ref{oneloopmass}) as a powerful check of our numerical integration algorithms by solving numerically equations (\ref{gluon_mass}), (\ref{dtGA}), and (\ref{dtGgh}) for small values of $g_k^2$.

\section{Renormalization schemes and the relation to the $\overline{\mathrm{MS}}$-scheme}
In order to give numbers in physical units we have to specify the renormalization-scheme.
For our purpose this specifies how a given numerical value for the dimensionless coupling $g_{k_0}$ is related to a scale $k_0$ in GeV units.
The numerical integration of the flow equations then yields all dimensionful quantities (as $q^2$ etc.) in units of $k_0$.
As emphasized in the introduction, the right hand side of the flow-equation (\ref{erge}) is both IR- and UV-finite for a suitable choice of the cutoff-function $R_k$ such as the one used in this work. Thus the formalism of exact flow-equations constitutes effectively a {\em{regularization}}-scheme.
In this context the {\em{renormalization}}-scheme is defined by fixing a starting value for the functional $\Gamma_k$ at a large scale $k=k_0$.
Physical quantities involving typical momenta $q^2 \ll k_0^2$ should only depend on a few relevant (or marginal) parameters in $\Gamma_{k_0}$ (up to corrections proportional to powers of $q^2/k_0^2$).
In this sense the solution of the exact flow equations must become independent of the renormalization scheme as $k \rightarrow 0$.
On the other hand, truncated versions of the flow equations will typically exhibit some dependence on the renormalization scheme.

In addition to the relation between $g_{k_0}$ and $k_0$ a complete determination of the renormalization scheme requires in our approximation also a specification of the initial values $z_A$ and $z_{gh}$ in equations (\ref{GA_1loop}) and (\ref{Ggh_1loop}).
We choose a parametrization
\bea
z_A&=&1 - \frac{3 \bar{g}^2}{16 \pi^2} \Biggl \lbrace \frac{95}{24}-\frac{{\mbox{Ei}}_1\!(1)}{2} + \frac{65}{24}{\mbox{Ei}}_1\!(1/2) + 2 e^{-1} - \frac{83}{12} e^{-1/2} \Biggr \rbrace - \delta_A \frac{\bar{g}^2}{16 \pi^2}
\label{zf}
\eea
and
\bea
z_{gh}&=&1 - \frac{3 \bar{g}^2}{16 \pi^2} \Biggl \lbrace \frac{5}{4}-\frac{{\mbox{Ei}}_1\!(1)}{4} + {\mbox{Ei}}_1\!(1/2) + \frac{1}{2} e^{-1} - 2 e^{-1/2} \Biggr \rbrace - \delta_{gh} \frac{\bar{g}^2}{16 \pi^2}
\label{zgh}
\eea
such that $\overline{G}_{A,k_0}(q^2 = k_0^2)  = G_{gh,k_0}(q^2 = k_0^2)  = k_0^2$ for $\delta_A = \delta_{gh} =0$.
In principle, the dependence of the final results on details of the relation between $g_{k_0}$ and $k_0$ or on $\delta$ can be used as a possible test of the truncation.
For $k \rightarrow 0$ the solution of the exact flow equations should not show any scheme dependence for the physical quantities.
A strong dependence of the approximative results on ``scheme parameters'' like $\delta$ would therefore indicate a shortcoming of the truncation.
(This is quite similar to the scheme dependence in a given order in a loop expansion.)
On the other hand, the choice of an ``optimal renormalization scheme'' may permit to keep the truncation errors on a minimal level.

The most important specification of the renormalization scheme concerns the value of $g_{k_0}$ at the initial scale $k_0$,
where we remind that $g_k$ is defined only implicitly.
It seems to be a good procedure to compute some physical quantity related to a mass scale in our scheme and to choose $g_{k_0}$ such that this quantity coincides with observation or the calculation in some other known scheme as the $\overline{\mathrm{MS}}$-scheme.
We only want to use perturbative physics as an input and would like a simple quantitative relation between our scheme and the $\overline{\mathrm{MS}}$-scheme.
A convenient physical quantity is the value of the Fourier transform of the effective heavy quark potential at a given momentum scale $V(q^2 = \mu^2)$.
The latter has been computed in the $\overline{\mathrm{MS}}$-scheme to two-loop order
\cite{MSbar2loop}
and one has
\bea
V_{\overline{\mathrm{MS}}}^{2-loop}(q^2) &=& \frac{4}{3} \frac{g_{\overline{MS}}^2(q^2)}{q^2} \Biggl \lbrace 1 + \left( \frac{31}{3} -\frac{10}{9} N_f \right) \frac{g_{\overline{MS}}^2(q^2)}{16 \pi^2} + \nonumber \\
&& \qquad \qquad \quad + \left[ \frac{4343}{18} + 54 \pi^2 - \frac{9 \pi^4}{4} + 66 \zeta(3) - \right. \nonumber \\
&& \qquad \qquad \quad  \left. - \left( \frac{1229}{27} + \frac{52}{3} \zeta(3) \right) N_f + \left( \frac{10}{9} N_f \right)^2 \right] \frac{g_{\overline{MS}}^4(q^2)}{(16 \pi^2)^2} \Biggr \rbrace 
\label{VMSbar}
\eea
The four quark interaction $\hat{V}(q^2)$ may differ from $V(q^2)$ by terms of order $g^4/q^2$ (see section 6) 
\bea
V(q^2) =\hat{V}(q^2) + \frac{\hat{\Delta}(q^2)}{q^2} = \tilde{V}(q^2) + \frac{\Delta(q^2)}{q^2}
\label{Vhat}
\eea
with $\Delta = \tilde{\Delta} + \hat{\Delta}$, cf.~equation (\ref{V}).
We therefore determine $g_{k_0}$ by imposing
\bea
V(\mu^2) &=& \frac{4}{3}  \frac{\bar{g}^2 \mu^4}{\overline{G}_A(\mu^2) G^2_{gh}(\mu^2)} + \frac{\Delta(\mu^2)}{\mu^2} \stackrel{!}{=} V_{\overline{\mathrm{MS}}}^{2-loop}(\mu^2) 
\label{Schemes}
\eea
for a fixed value of $\mu$ in the perturbative range.

We still need the value of $\Delta(\mu^2)$.
To lowest order it can be inferred by comparing a one loop calculation of $\tilde{V}(q^2)$ in the $\overline{\mathrm{MS}}$-scheme with eq.~(\ref{VMSbar}).
In one loop order the difference $V-\tilde{V}$ is ultraviolet finite in Landau gauge.
One obtains
\bea
\Delta(\mu^2) = - 5 \frac{g^4_{\overline{\mathrm{MS}}}(\mu^2)}{16 \pi^2}
\label{Delta}
\eea
and the relations (\ref{Schemes}) and (\ref{Delta}) are now sufficient to express $\bar{g}^2 \equiv g_{k_0}^2$ as a function of $\delta$ and $g^2_{\overline{\mathrm{MS}}}(\mu)$. 
The nontrivial part of the solutions of flow equations concerns then the shape of $\tilde{V}(q^2)$ or $V(q^2)$ for $q^2 \neq \mu^2$.

Let us next turn to the choice of an optimal renormalization scheme which minimizes the importance of neglected corrections.
For the choice of $\bar{g}$ (or $\mu$ in equation (\ref{Schemes})) we rely on the fact that $V$ is two loop exact for $q^2 = \mu^2$.
All truncation errors in order $g^6$ therefore only affect the behaviour of $V$ for $q^2 \neq \mu^2$.
Since we are mainly interested in low values of $q^2$ we will choose $\mu^2$ as low as possible with the constraint that the two loop approximation $V_{\overline{\mathrm{MS}}}^{2-loop}$ should still remain valid.
As a typical scale we will take $\mu = 3 \Lambda_{\overline{\mathrm{MS}}}$ with $\Lambda_{\overline{\mathrm{MS}}}$ the two loop confinement scale in the ${\overline{\mathrm{MS}}}$-scheme ($N_f=0$).
For an optimal choice of $\delta$ we will further assume (see section 6) that the difference $\Delta(q^2) - \Delta(\mu^2)$ remains small as compared to $q^2 \tilde{V}(q^2)$ for a certain momentum range $q^2$ around $\mu^2$.
This will allow us to relate the momentum dependence of $\tilde{V}(q^2)$ and $V(q^2)$.
By expanding equations (\ref{V}), (\ref{GA_1loop}) and (\ref{Ggh_1loop}) for large $q^2$ in powers of $g^2$ it is easy to convince oneself that the momentum dependence of $q^2 \tilde{V}(q^2)$ is influenced in two loop order by the value of $\delta$.
With our approximations this also extends to $q^2 V(q^2)$.
We will adopt a choice of $\delta$ for which the deviation between $V(q^2)$ and $V_{\overline{\mathrm{MS}}}^{2-loop}(q^2)$ remains small for a certain range $q^2 > \mu^2$.
(Of course, a truncation which is exact in two loop order would fix the two loop momentum dependence unambiguously and $V$ could deviate from $V_{\overline{\mathrm{MS}}}^{2-loop}$ only in higher loop order.
Our optimal choice of $\delta$ partly anticipates these higher order effects by minimizing their role in a whole range $q^2 \geq \mu^2$.)
For the practical calculations with $N_f = 0$ we will take $\delta_A = \delta_{gh} = \delta = -3$.
In the following we will always work with this ``optimal renormalization scheme''.
This turns the disadvantage of the scheme ambiguity into an advantage by making use of independent perturbative calculations for $q^2 \geq \mu^2$.
Of course, this optimal scheme does not tell us what will happen for $q^2 < \mu^2$ and it is in this region where the solution of the flow equation should extend our information beyond perturbation theory.

The choice of the scheme also affects the size of the neglected corrections $\sim \Delta(q^2) - \Delta(\mu^2)$ in the relation between $\tilde{V}$ and $V$.
The fact that this quantity vanishes identically in one loop order and is almost zero for a whole range $q^2 \geq \mu^2$ strongly suggests that this quantity is indeed small compared to $q^2 \tilde{V}(q^2)$ in the optimal renormalization scheme, at least for a certain range $q^2 < \mu^2$.
We thus expect $\tilde{V}(q^2)+\frac{\Delta(\mu^2)}{q^2}$ to be a good approximation to the heavy quark potential in this range.
On the other hand, the information on $\tilde{V}(q^2)$ can be transfered to a precise determination of the four quark interaction $\hat{V}(q^2)$ only once the term $\tilde{\Delta}$ in equation (\ref{V}) is computed.
For the time being we identify $\hat{V}$ with $V$ (i.e.~we assume $\tilde{\Delta} = - \frac{5 \bar{g}^4}{16 \pi^2}$).

\begin{figure}[H]
\begin{center}
\epsfig{file=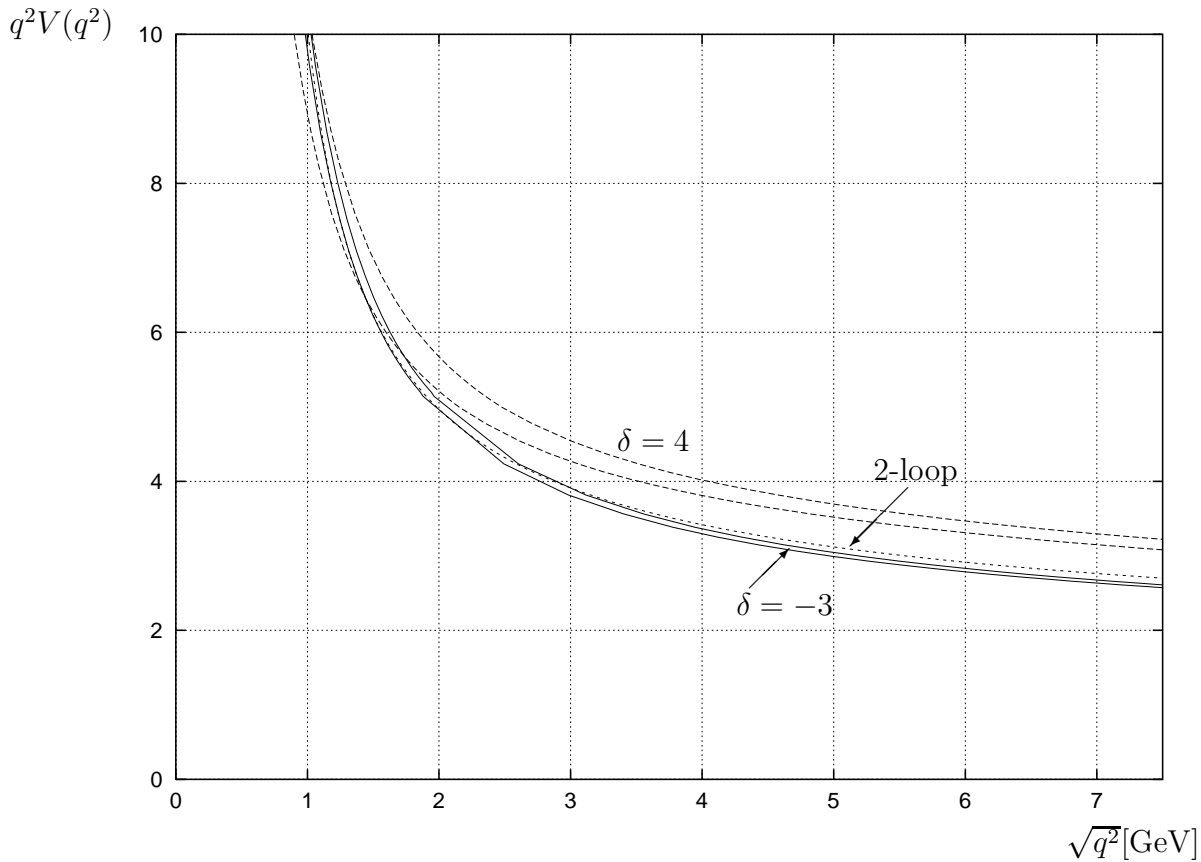,width=15cm}
\put (-450,295) {$q^2 V(q^2)$}
\put (-50,-15) {$\sqrt{q^2} [\mathrm{GeV}]$}
\put (-123,126) {2-loop}
\put (-110,123) {\vector(-1,-1){22}}
\put (-220,137) {$\delta=4$}
\put (-175,75) {$\delta=-3$}
\put (-170,85) {\vector(1,1){15}}\\
\end{center}
\renewcommand{\baselinestretch}{1.2}\normalsize
\caption{Comparison of $q^2 V(q^2)$ and $q^2 V^{2-loop}_{\overline{\mathrm{MS}}}(q^2)$ for different values of $\delta$ and $\mu$ (see equations (22), (23) and (26)).
The dotted line gives the two loop ${\overline{\mathrm{MS}}}$-result, the solid lines correspond to $\delta=-3$ and the dashed lines are obtained for $\delta=4$.
The scale $\mu$ is choosen to be $3 \Lambda_{\overline{\mathrm{MS}}}$ for the lower (upper) solid (dashed) line and $4.5 \Lambda_{\overline{\mathrm{MS}}}$ for the upper (lower) solid (dashed) line.}
\renewcommand{\baselinestretch}{1.7}\normalsize
\end{figure}
Results for $q^2 V(q^2)$ with two values $\mu_1 = 3 \Lambda_{\overline{\mathrm{MS}}}$ and $\mu_2 = 4.5 \Lambda_{\overline{\mathrm{MS}}}$ are shown in figure 3. 
(For the choice of the momentum scale see section 6).
Since the present truncation of the renormalization group equations reproduces the running of $q^2 V(q^2)$ only to one loop accuracy, the scheme dependence can still be relatively important, as can be seen by comparing our results for the two values of $\mu$ but with the same choice of $\delta=4$.
For a fixed value of $\mu$ we should, following the discussion above, find the value of $\delta$ such that the two loop result is approximately reproduced in a range $q^2 \geq \mu^2$.
Varying $\mu$ within this range then does not change the results considerably anymore, as can be seen from the curves with $\delta = -3$.

\section{Effective heavy quark potential}
It is tempting to associate $\tilde{V}(q^2) - \frac{5 \bar{g}^4}{16 \pi^2 q^2}$ with the Fourier transform of the heavy quark potential $V(q^2)$.
(See 
\cite{UliManfredAxel}
for an identification of $\tilde{V}(q^2)$ with $V(q^2)$.)
The potential can then be inferred from the gluon and ghost propagators for $k \rightarrow 0$ as
\bea
V(q^2) = \frac{4}{3} \frac{\bar{g}^2 q^4}{{\overline{G}}_A(q^2) G_{gh}(q^2)} - \frac{5 \bar{g}^4}{16 \pi^2 q^2}
\label{Potential}
\eea
As we have argued in the preceeding section, we expect this to be a valid approximation in Landau gauge and for the ``optimal renormalization scheme'' used here.
On the other hand, this identification is certainly not valid for general gauges and renormalization schemes.
It is important to keep in mind the structural differences between the four quark interaction $\hat{V}$ and the heavy quark potential $V$.
First of all, we observe that $V$ is a gauge invariant quantity whereas $\hat{V}$ and $\tilde{V}$ are not.
For example, the momentum dependence of $q^2 \tilde{V}(q^2)$ coincides with the one of $q^2 V(q^2)$ in one loop order only in Landau gauge.

The static potential is commonly defined in terms of Wegner-Wilson-Loops
\cite{Fischler}.
An expansion of the expectation value of the Wegner-Wilson loop (or a similar gauge invariant quantity involving four quarks) in terms of Green's functions for quarks and gluons not only contains a lowest order four-quark interaction (\ref{quark_action}) but also higher terms $\propto g <(\bar{\psi}\psi)^2 A>$, $g^2 <(\bar{\psi}\psi)^2 A^2>$ etc..
After solving the field equations for $A$ such terms would appear in the form of effective interactions involving six or more quarks.
They contribute to $V-\hat{V}$ in higher orders in $g$.

Alternatively, one may think of using the effective four quark interaction for a computation of ``heavy quark scattering'' in the form of some physical and therefore gauge invariant cross section.
However, any reasonable definition of heavy quark scattering involves (almost) on shell quarks and must include soft or collinear gluons.
In contrast, $\hat{V}(q^2)$ is evaluated for Euclidean quark momenta and nonvanishing infrared cutoff $k_F$.
The relative suppression of quark-vertex corrections ($Y_\psi$), box diagrams ($E$) and quark wave function effects ($Z_\psi-1$) for low $q^2$ is due to the infrared cutoff $k_F$. 
It may not be the same for on-shell massive quarks with $k_F=0$ 
\cite{ChQCD}.
(Only for Euclidean momenta we are guaranteed that quark propagators always give suppression factors $\propto k_F^{-1}$ or $m_H^{-1}$ with $m_H$ the heavy quark mass if the diagram is ultraviolet finite.)

In summary, we conclude that on the conceptual level the difference between $\tilde{V}$ and $V$ is far from being trivial.
Nevertheless, we know that for Landau gauge fixing the difference $q^2\left( V(q^2) - \tilde{V}(q^2) \right) = \Delta(q^2)$ can only contain terms $\propto g^4$, $g^6 \ln q^2 / \mu^2$ etc. in a perturbative expansion.
The constant term $\propto g^4$ is incorporated in our approximation (\ref{Delta}).
The optimal renormalization scheme minimizes the combined error from the neglection of the $q^2$-dependence of $\Delta$ and the truncation in the computation of $\tilde{V}$.
Only in this scheme the neglected corrections to $V$ remain small for a whole range $q^2 \geq \mu^2$, and, by continuity, also for a certain range $q^2 < \mu^2$.
It seems then reasonable to propose that the solution of the flow equation yields in the present setting a reasonable picture of $V(q^2)$ down to relatively low values of $q^2$, say $\sqrt{q^2} \approx 300$~MeV.

In order to test the approximations we will compare our results for $V(q^2)$ with phenomenological quark potentials used for charmonium, as for example the Richardson potential
\cite{Richardson}
\bea
V_R(q^2) &=& \frac{64 \pi^2}{27} \frac{1}{q^2 \ln \left(1 + \frac{q^2}{\Lambda_R^2} \right)} \qquad , \qquad \Lambda_R = 0.4~{\mathrm{GeV}}
\label{Richardson}
\eea
Of course, such a comparison would be more meaningful if the effects of light quark flavors were included.
Nevertheless, results of a ``quenched approximation'' ($N_f=0$) should not come out too far from reality provided the overall mass scale is properly chosen.

In this section we will present the results of a numerical integration of the system of integro-differential equations (\ref{dtGA}) and (\ref{dtGgh}) with a gluon-mass counterterm given by (\ref{gluon_mass}) and a running coupling as discussed in section 3.
Since this amounts to a calculation without light quarks ($N_f=0$) we compare our results to perturbation theory in a pure Yang-Mills theory.
All mass and momentum scales are expressed in terms of the two loop confinement scale $\Lambda_{\overline{\mathrm{MS}}}$ for $N_f=0$.
The value of $\Lambda_{\overline{\mathrm{MS}}}$ in physical units will depend on the problem under consideration.
We concentrate in this section on a computation of the heavy quark potential in the quenched approximation.
In order to relate $\Lambda_{\overline{\mathrm{MS}}}$ to the real world where $N_f=3$ we use the condition that the ${\overline{\mathrm{MS}}}$ gauge coupling for $N_f=0$ should coincide with the one for $N_f=3$ at the edge of the range of validity of perturbation theory.
(This minimizes trivial effects from the different running of the coupling with and without light quarks in the perturbative range.)
In practice, we identify the $N_f=0$ gauge coupling with the ``real'' gauge coupling at a scale where $\beta_{g^2} / g^2 = -1.5$, resulting in $\Lambda_{\overline{\mathrm{MS}}} = 315$~MeV (for a three-flavour confinement scale $\Lambda_{\overline{\mathrm{MS}}}^{(N_f=3)} = 285$~MeV).
For a computation of the effective quark interaction discussed in the introduction the choice of $\Lambda_{\overline{\mathrm{MS}}}$ will be different.
As discussed in the conclusions, the $N_f=0$ gauge coupling should then coincide with the ``real'' $N_f=3$ coupling at the scale $k_\psi$.

We start the numerical integration of the flow equations at $k_0 = 20 \Lambda_{\overline{\mathrm{MS}}} = 6.3$~GeV and use $\mu = 3 \Lambda_{\overline{\mathrm{MS}}} = 945$~GeV and $\delta = -3$ for the definition of the renormalization scheme in equation (\ref{Schemes}).
For the infrared value of the gauge coupling we take $\alpha_\star = 1.5$.
We will first discuss the results for the heavy quark potential $V(q^2)$ as defined in (\ref{Potential}).
In figure 2 we have displayed $q^2 V(q^2)$ as a function of $\sqrt{q^2}$ for various values of the infrared cutoff $k$.
For $k= 500$~MeV the effects of fluctuations are still moderate since the low momentum fluctuations are suppressed.
The value of $q^2 V(q^2)$ remains near the classical value $\frac{4}{3} g^2(\mu) \sim 11$.
As $k$ is lowered to $100$~MeV, additional fluctuations are included.
One observes a strong rise of $q^2 V(q^2)$ for small $q^2$.
Finally, for $k \rightarrow 0$ all fluctuations are included and the solid curve gives our estimate for the effective potential.
It is this latter curve which has been plotted for different values of $\mu$ and $\delta$ and compared with perturbation theory in figure 3.
For nonzero $k$ we note that $V_k(q^2)$ gives a good approximation to $V_0(q^2)$ for $\sqrt{q^2} \grgl 3 k$.
Otherwise speaking, the infrared cutoff substantially influences the potential only in the region $\sqrt{q^2} \klgl 3 k$.
This is a more quantitative version of our statement that fluctuations with momenta $q^2 \klgl k^2$ have only little influence on the $n$-point functions at large $q^2$.
In figure 4 we compare the final value of $V(q^2)$ with the Richardson potential for three values $\alpha_\star = 1,\, 1.5$ and $2$.
The agreement for $\alpha_\star = 1$ -- $1.5$ is quite remarkable.
Overall, we find that our results smoothly interpolate between the two loop potential for large $q^2$ and phenomenologically acceptable potentials for $300~{\mathrm{MeV}} \klgl \sqrt{q^2} \klgl 800~{\mathrm{MeV}}$.
\begin{figure}[H]
\begin{center}
\epsfig{file=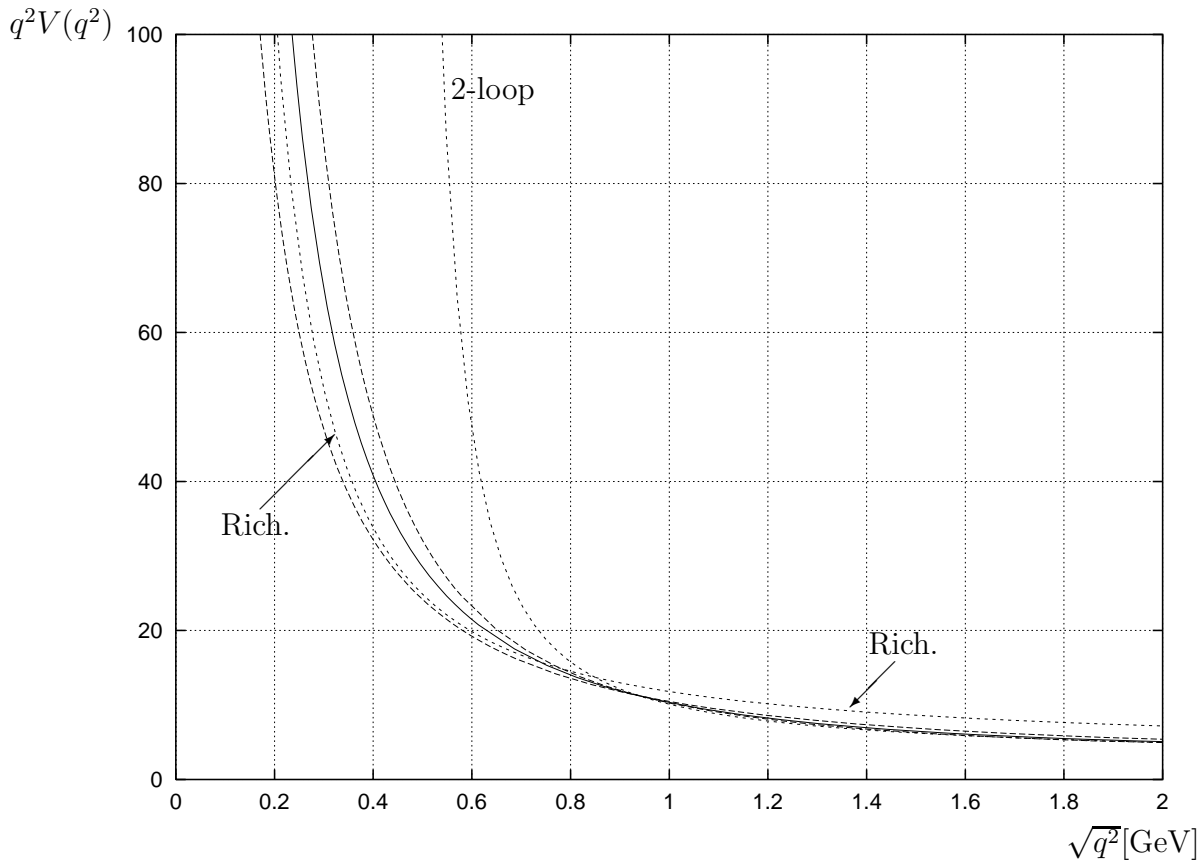,width=15cm}
\put (-450,295) {$q^2 V(q^2)$}
\put (-50,-15) {$\sqrt{q^2} [\mathrm{GeV}]$}
\put (-283,270) {2-loop}
\put (-370,105) {Rich.}
\put (-355,115) {\vector(1,1){28}}
\put (-125,60) {Rich.}
\put (-115,57) {\vector(-1,-1){17}}\\
\end{center}
\renewcommand{\baselinestretch}{1.2}\normalsize
\caption{Heavy quark potential for different values of $\alpha_\star$.
The solid line corresponds to $\alpha_\star = 1.5$ whereas the dashed lines are for $\alpha_\star = 1$ (lower dashed line) and $\alpha_\star = 2$ (upper dashed line).
Also shown are the two loop potential and the Richardson fit as dotted lines.}
\renewcommand{\baselinestretch}{1.7}\normalsize
\end{figure}

For a more detailed discussion of the momentum dependence of $V(q^2)$ it is useful to define the quantity 
\bea
\chi_V(q^2)&=& 2 \frac{\partial}{\partial \ln q^2} \ln \left( q^2 V(q^2) \right)
\label{chiV}
\eea
We have plotted this quantity in figure 5, again for different values of $k$.
In the limit of a constant $\chi_V$ one has 
\bea
\chi_V = \eta_V \quad \Longrightarrow \quad V(q^2) = \frac{A}{(q^2)^{1-\frac{\eta_V}{2}}}
\label{chiVexample}
\eea
\begin{figure}[H]
\begin{center}
\epsfig{file=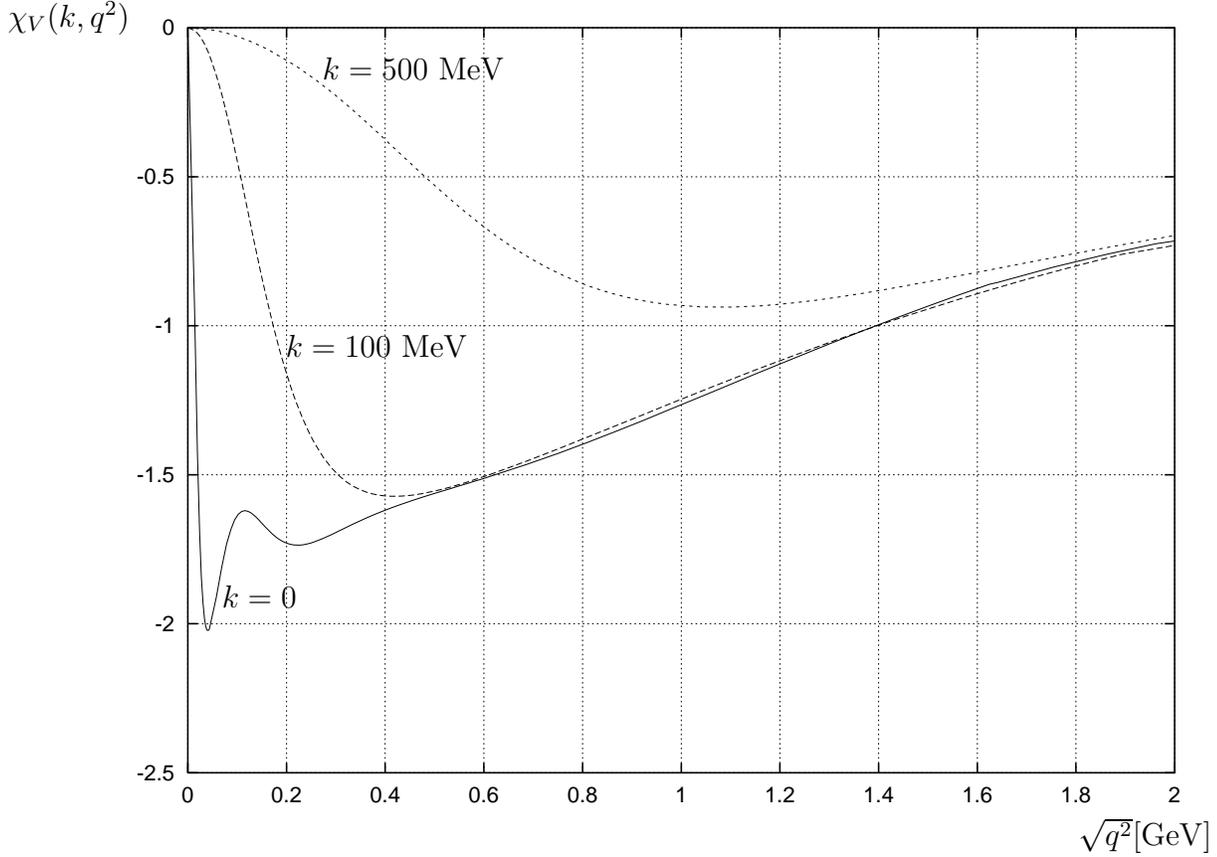,width=15cm}
\put (-455,295) {$\chi_V(k,q^2)$}
\put (-50,-15) {$\sqrt{q^2} [\mathrm{GeV}]$}
\put (-374,75) {$k=0$}
\put (-336,275) {$k=500$~MeV}
\put (-350,170) {$k=100$~MeV}\\
\end{center}
\renewcommand{\baselinestretch}{1.2}\normalsize
\caption{Momentum dependent anomalous dimension $\chi_V(q^2)$ for different values of the cutoff $k$.
The dotted line corresponds to $k = 500$~MeV, the dashed line is at $k = 100$~MeV, and the solid line displays the final result for $k=0$.}
\renewcommand{\baselinestretch}{1.7}\normalsize
\end{figure}

\noindent A potential $V(q^2) \propto q^{-4}$ which is the Fourier transform of a linear rising confinement potential corresponds to $\chi_V(q^2 \rightarrow 0) = -2$.
Our results have a tendency to approach this value.
For $\sqrt{q^2} \klgl 400$~MeV the dependence on the truncation (i.e.~the value of $\alpha_\star$ in equation (\ref{IRalpha})) becomes important, however.
It is apparent that the truncation uncertainty affects a derivative quantity as $\chi_V$ much stronger than the potential itself.
In figure 6 we display the function $\chi_V(q^2)$ for $k=0$ for different values of $\alpha_\star$.

We observe that $\chi_V$ is directly related to the $\beta$-function for a suitably defined running coupling
\bea
g_V^2(q^2) &=& \frac{3}{4} q^2 V(q^2) \nonumber \\
\frac{\partial}{\partial \ln \sqrt{q^2}} g_V^2 &=& \beta_{g^2_V}(g_V^2) = \chi_V g_V^2
\label{gV}
\eea
\begin{figure}[H]
\begin{center}
\epsfig{file=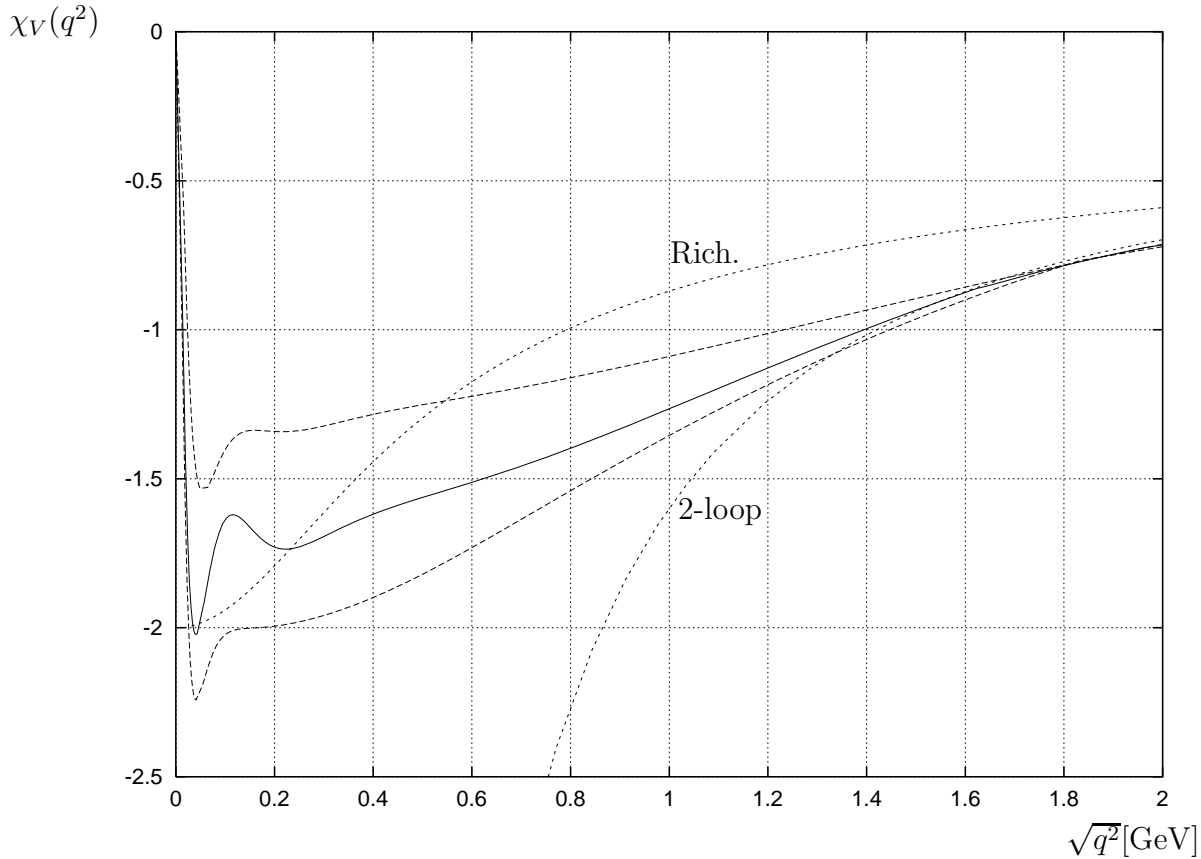,width=15cm}
\put (-450,295) {$\chi_V(q^2)$}
\put (-50,-15) {$\sqrt{q^2} [\mathrm{GeV}]$}
\put (-197,110) {2-loop}
\put (-200,207) {Rich.}\\
\end{center}
\renewcommand{\baselinestretch}{1.2}\normalsize
\caption{$\chi_V(q^2)$, for $k=0$ and different values of $\alpha_\star$: the solid line is for $\alpha_\star = 1.5$, the dashed lines are for $\alpha_\star = 1$ (upper dashed line) and $\alpha_\star = 2$ (lower dashed line).
The dotted lines give the two loop value and the result corresponding to the Richardson potential.}
\renewcommand{\baselinestretch}{1.7}\normalsize
\end{figure}

\noindent The two-loop perturbative value is given by the universal two-loop $\beta$-function and thus reads
\bea
\chi_V(q^2) = - 22 \frac{g_V^2(q^2)}{16 \pi^2} - 204 \frac{g_V^4(q^2)}{(16 \pi^2)^2}
\label{chiVpert}
\eea
Extracting $g_V^2(q^2)$ from our numerical solution we have also plotted the function (\ref{chiVpert}) in figure 6.
The agreement of our numerical solution for $\chi_V$ with this curve for high $q^2$ is remarkable. 
We emphasize that the two loop running of $g_k$ in (\ref{beta_g}) does not automatically imply a two loop running of the momentum dependent $g_V(q^2)$ for large $q^2$.
At the present level of accuracy it is impossible to distinguish which part of the deviation of $\chi_V(q^2)$ from (\ref{chiVpert}) is due to shortcomings of the truncation or to the difference between the four quark interaction $\tilde{V}(q^2)$ and the heavy quark potential.

For small momenta, the perturbative result for $\chi_V$ is no longer reliable.
For the Richardson potential  (equation (\ref{Richardson})), the behaviour of $\chi_V$ reads
\bea
\chi_V^{Rich}(q^2) = -2 \frac{q^2}{q^2 + \Lambda_R^2} \frac{1}{\ln \left( 1 + \frac{q^2}{\Lambda_R^2} \right) }
\label{chiRich}
\eea
This curve is also displayed in figure 6.
Again, we observe the interpolation between the two loop and Richardson potential.

The quantity $g_V^2$ may also be used in order to parametrize our results for $V(q^2)$ in a form which allows for simple rescaling of the overall mass unit.
Let us introduce the dimensionless variable $z = q^2 / \Lambda_{\overline{\mathrm{MS}}}^2$.
Our results for the four quark interaction can be summarized in the fit ($\alpha_\star = 1.5$) 
\bea
\frac{g_V^2}{4 \pi} (z) &=& \frac{\gamma(z)}{4\pi} \Biggl \lbrace 1 + \frac{31}{3} \frac{\gamma(z)}{16 \pi^2} + \left[\frac{4343}{18} + 54 \pi^2 - \frac{9 \pi^4}{4} + 66 \zeta (3) \right] \frac{\gamma^2(z)}{\left(16 \pi^2 \right)^2} \Biggr \rbrace \times \nonumber \\
&& \times \Biggl \lbrace 0.977 + \frac{3.283}{1+1.154 z} + \frac{7.262}{\left( 1+48.24 z\right)^2} - \frac{3.584}{\left(1+0.269 z\right)^3} \Biggr \rbrace
\label{fit}
\eea
where
\bea
\gamma(z) = \frac{48 \pi^2}{33} \left\{ 1 - \frac{102}{121} \frac{\ln \left( \ln(z+3.507) \right)}{\ln(z+3.507)} \right\} \frac{1}{\ln(z+3.507)}
\label{fit2}
\eea
This fit reproduces our results in the range $0.05 \klgl z \klgl 100$ with a relative error $< 0.5 \%$. 

Since the agreement with phenomenological potentials is satisfactory in the range $300~{\mathrm{MeV}} < \sqrt{q^2} < 800~{\mathrm{MeV}}$ the charmonium levels computed from our approximation to $V(q^2)$ will be close to the experimental values.
One may ask if a string tension $\sigma$ can also be extracted.
Instead of determining $\sigma$ from the behaviour of $V(q^2)$ for $q^2 \rightarrow 0$ we will use here a more phenomenological definition by fitting the quantity $\sigma(q^2) = \frac{3}{16 \pi} q^4 V(q^2)$ for $\frac{1}{2} \Lambda_{\overline{\mathrm{MS}}} < \sqrt{q^2} < \frac{3}{4} \Lambda_{\overline{\mathrm{MS}}}$ as
\bea
\sigma(q^2) = \frac{3}{16 \pi} q^4 V(q^2) = \sigma + c q^2
\label{fitsigma}
\eea
In figure 7 we display $\sigma(q^2)$ for different values of $\alpha_\star$. 
For a range $\alpha_\star = 1 - 2$ we extract
\bea
0.13~{\mathrm{GeV}}^2 < \sigma < 0.46~{\mathrm{GeV}}^2 
\label{ressigma}
\eea
\begin{figure}[H]
\begin{center}
\epsfig{file=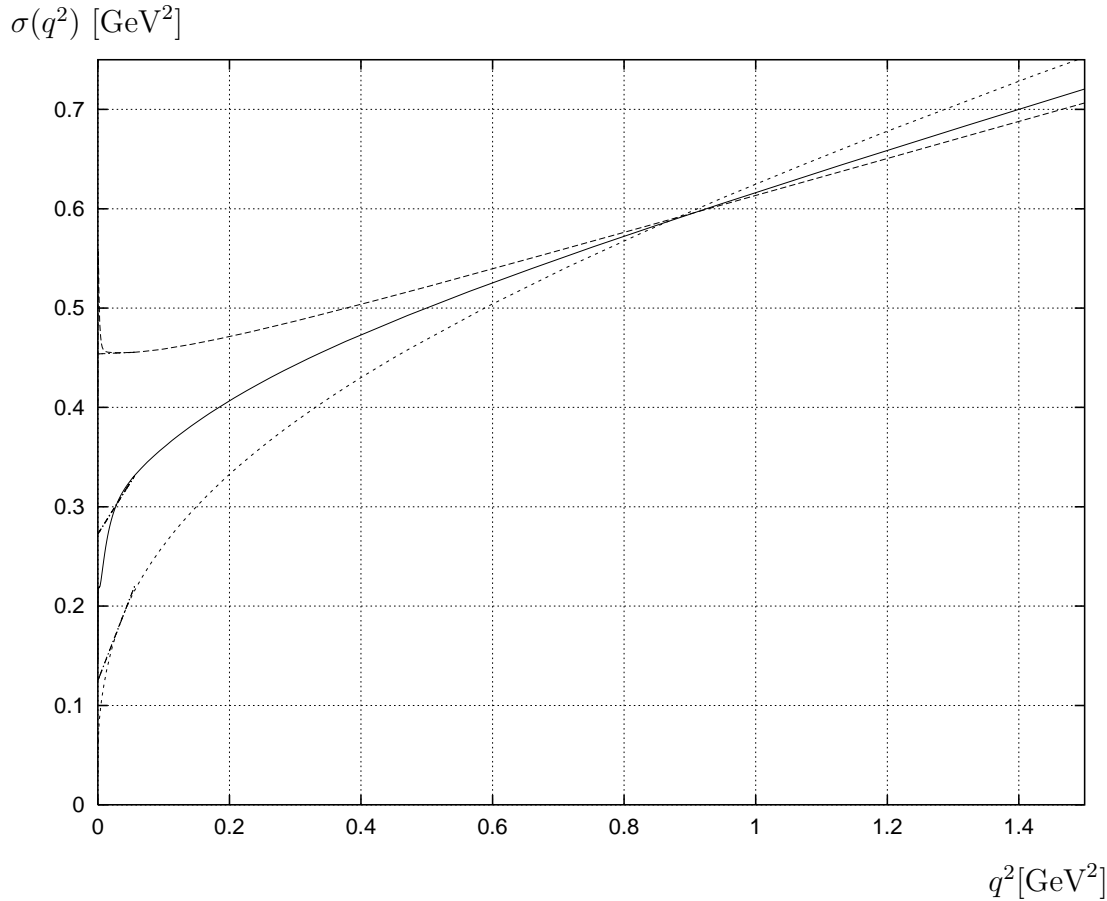,width=15cm}
\put (-420,305) {$\sigma(q^2)~[\mathrm{GeV}^2]$}
\put (-50,-20) {$q^2 [\mathrm{GeV}^2]$}\\
\end{center}
\renewcommand{\baselinestretch}{1.2}\normalsize
\caption{String tension: The plot shows $\sigma(q^2) = \frac{3}{16 \pi} q^4 V(q^2)$ for different values of $\alpha_\star$.
The solid line is for $\alpha_\star = 1.5$, the dotted line is for $\alpha_\star = 1$ and the dashed line gives the result for $\alpha_\star = 2$.
We also indicate the straight line extrapolations from the region $0.025~{\mathrm{GeV}}^2 < q^2 < 0.06~{\mathrm{GeV}}^2$.}
\renewcommand{\baselinestretch}{1.7}\normalsize
\end{figure}

\noindent and $1.7 >  c > 0.03$.
In view of the uncertainties and missing effects of light quark fluctuations the agreement with the observations from quarkonia ($\sigma \approx 0.18~{\mathrm{GeV}}^2$) seems reasonable.
Obviously, our approximations are insufficient to extrapolate $V(q^2)$ to $q^2 \rightarrow 0$ and to establish a ``confinement behaviour'' $V \sim q^{-4}$ unambiguously.
This is apparent from the qualitative change of $q^4 V(q^2)$ in dependence on $\alpha_\star$ (figure 7).
We observe a bifurcation in the sense that for $\alpha_\star > \alpha_\star^{(c)}$ the limit of $q^4 V(q^2)$ diverges for $q^2 \rightarrow 0$ whereas for $\alpha_\star < \alpha_\star^{(c)}$ it vanishes.
Only for $\alpha_\star = \alpha_\star^{(c)} \sim 2$ the potential scales exactly $\propto q^{-4}$.
One may speculate that a proper treatment of the flow of the effective coupling could lead to a combined fixed point for $g_k$ (corresponding to $\alpha_\star^{(c)}$), $\frac{\partial Z_F}{\partial q^2}(q^2=0)$ and $Y_F(q^2 = 0)$, whereas $Z_F(q^2=0)$ vanishes for $k \rightarrow 0$.
It would be interesting to investigate the scaling behaviour of the flow equation for the ansatz (\ref{first_truncation}) with $Z_F(q^2) = K_F q^2$ and $Y_F$, $K_F$ independent of $q^2$.
The vertices should then be defined by this truncation and the effective coupling $g_k$ could be extracted a posteriori.
\begin{figure}
\begin{center}
\epsfig{file=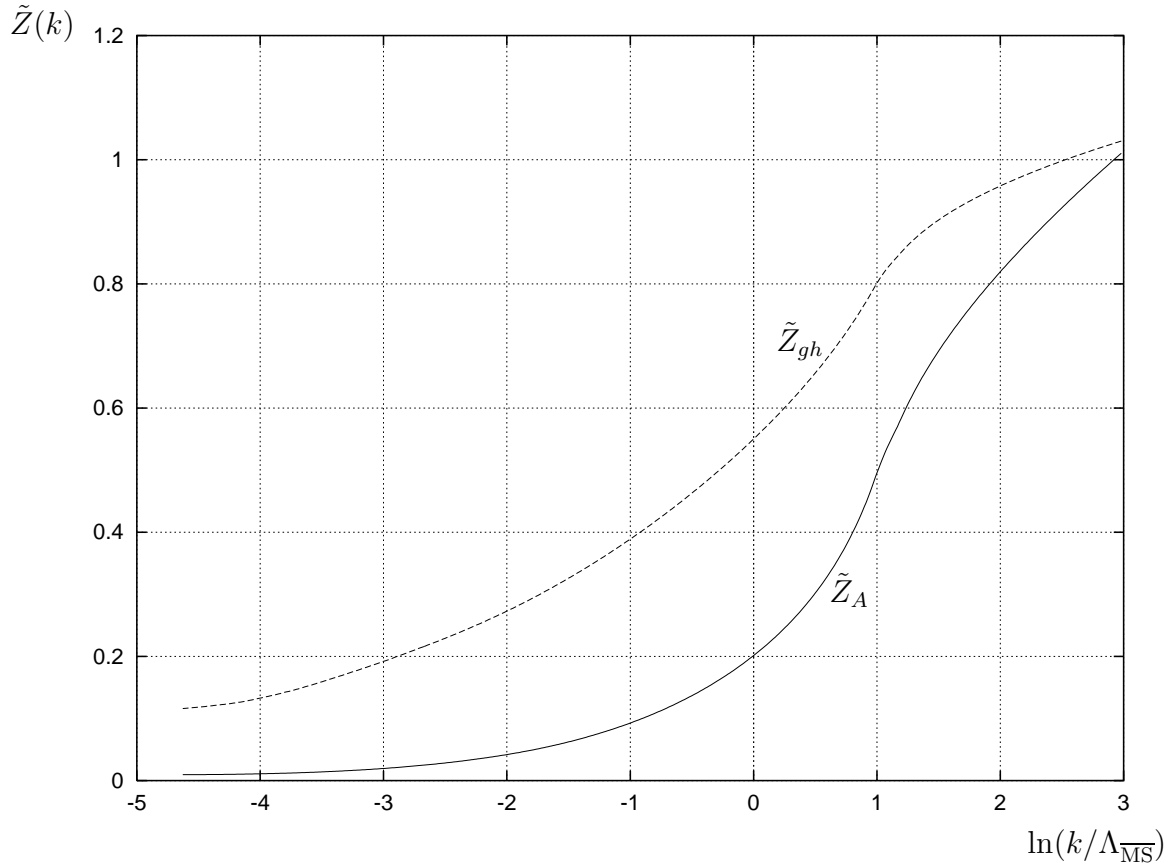,width=15cm}
\put (-435,295) {$\tilde{Z}(k)$}
\put (-50,-15) {$\ln (k / \Lambda_{\overline{\mathrm{MS}}})$}
\put (-125,80) {$\tilde{Z}_A$}
\put (-145,175) {$\tilde{Z}_{gh}$}\\
\end{center}
\renewcommand{\baselinestretch}{1.2}\normalsize
\caption{Wave function renormalizations $\tilde{Z}_A$ (solid line) and $\tilde{Z}_{gh}$ (dashed line) as functions of $\ln (k / \Lambda_{\overline{\mathrm{MS}}})$, for $\alpha_\star = 1.5$.}
\renewcommand{\baselinestretch}{1.7}\normalsize
\end{figure}

Finally, it is also worthwhile to study the momentum and scale dependence of the gluon and ghost propagators in Landau gauge.
The overall size of $\overline{G}_A$ and $G_{gh}$ is given by the wave function renormalizations $\tilde{Z}_A$ and $\tilde{Z}_{gh}$ (equations (\ref{WFRs})).
The $k$-dependence of $\tilde{Z}(k)$ is plotted in figure 8 and we observe in particular a strong decrease of $\tilde{Z}_A$ for $k \rightarrow 0$.
The momentum dependence of the propagators is shown in figure 9.
Our results seem to indicate that $\overline{G}_A(q^2)/q^2$ vanishes for $q^2 \rightarrow 0$.
\begin{figure}[H]
\begin{center}
\epsfig{file=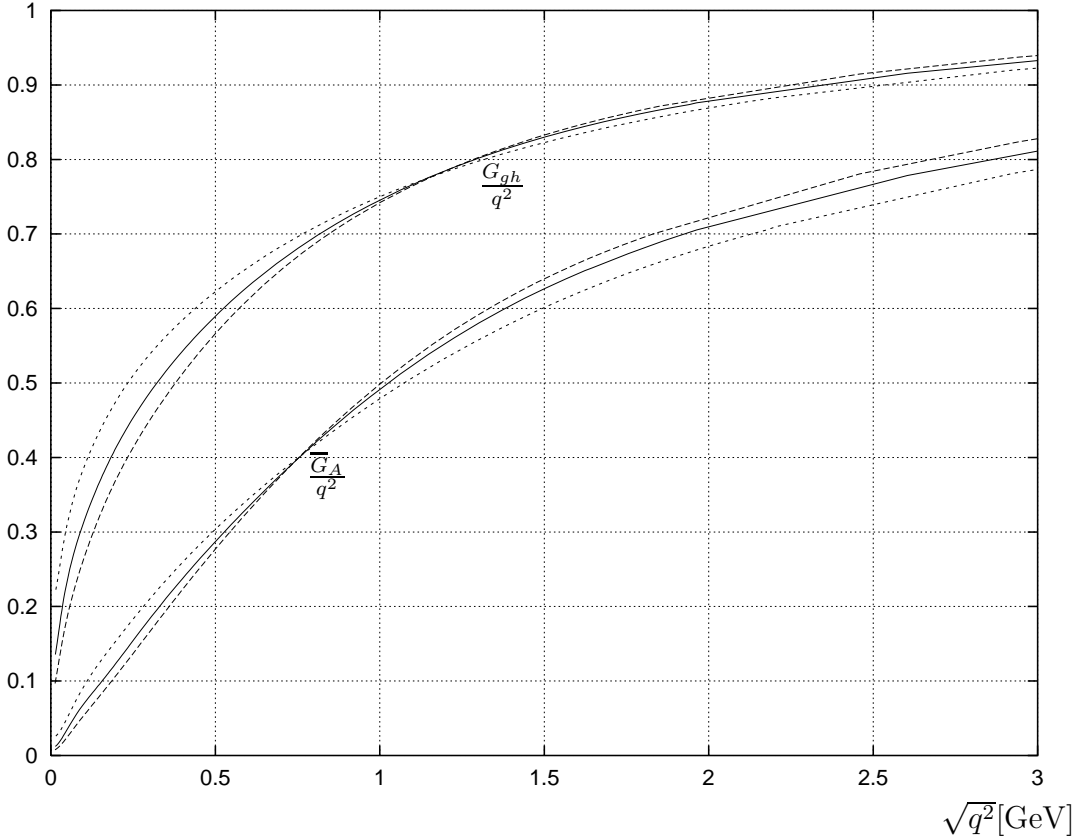,width=15cm}
\put (-50,-15) {$\sqrt{q^2} [\mathrm{GeV}]$}
\put (-290,115) {$\frac{\overline{G}_A}{q^2}$}
\put (-225,225) {$\frac{G_{gh}}{q^2}$}\\
\end{center}
\renewcommand{\baselinestretch}{1.2}\normalsize
\caption{$\overline{G}_A(q^2)/q^2$ (lower lines) and $G_{gh}(q^2)/q^2$ (upper lines), for $k=0$ and different values of $\alpha_\star$: the solid lines are for $\alpha_\star = 1.5$, the dotted lines correspond to $\alpha_\star = 1$ and the dashed lines give the results for $\alpha_\star = 2$.}
\renewcommand{\baselinestretch}{1.7}\normalsize
\end{figure}

For a more detailed investigation of the momentum dependence we introduce ``momentum dependent anomalous dimensions''
\bea
\chi_i(q^2,k) = - 2 \frac{\partial}{\partial \ln q^2} \ln \frac{G_i(q^2)-G_i(0)}{\tilde{Z}_i q^2}
\label{chi}
\eea
The quantity $\chi_V$ in equation (\ref{chiV}) is related to $\chi_A$ and $\chi_{gh}$ by
\bea
\chi_V(q^2)&=& \left( \chi_A(q^2) + 2 \chi_{gh}(q^2) \right) \left( 1 - \frac{\Delta}{q^2 V(q^2)} \right)
\label{chiVandchifields}
\eea
where we observe that the correction $\propto \Delta$ becomes rapidly insignificant for small $q^2$.
We have plotted $\chi_A$ and $\chi_{gh}$ in figure 10.
For large $q^2$ these quantities should approach the one loop anomalous dimensions for the wave function renormalization of the gluon and ghost fields,
\begin{figure}[H]
\begin{center}
\epsfig{file=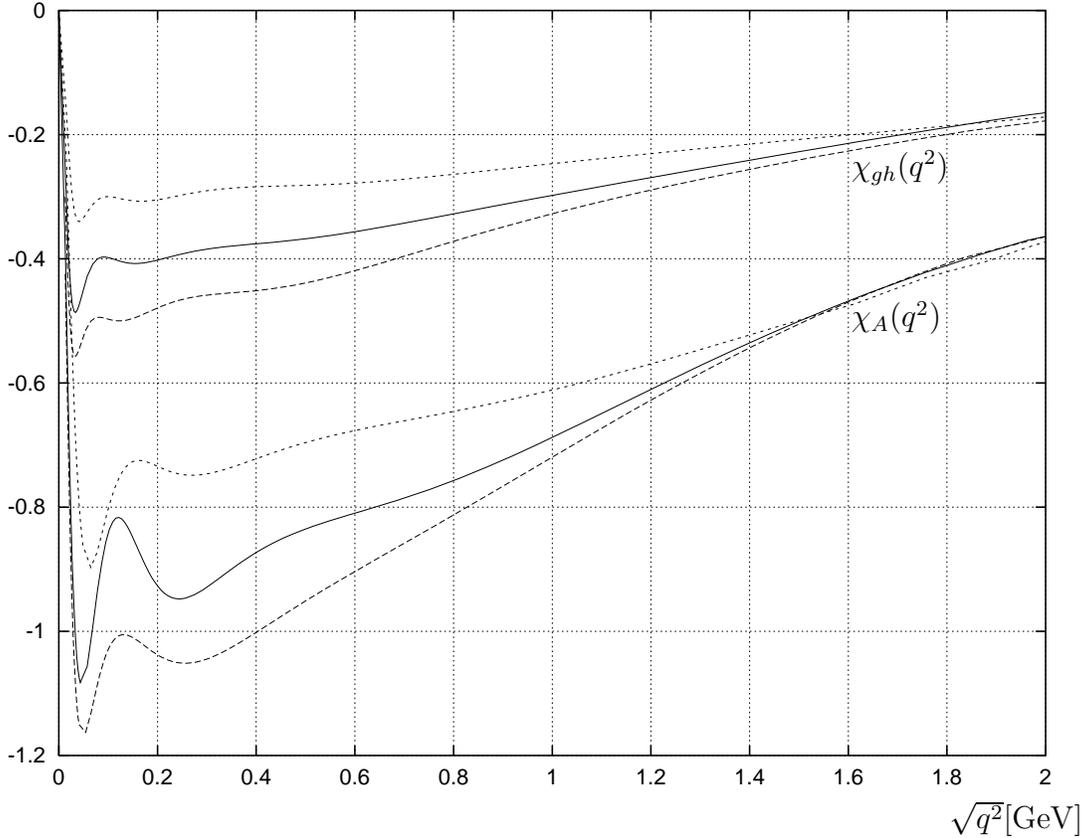,width=15cm}
\put (-50,-15) {$\sqrt{q^2} [\mathrm{GeV}]$}
\put (-87,175) {$\chi_A(q^2)$}
\put (-87,232) {$\chi_{gh}(q^2)$}\\
\end{center}
\renewcommand{\baselinestretch}{1.2}\normalsize
\caption{Momentum dependent anomalous dimensions $\chi_A(q^2)$ (lower lines) and $\chi_{gh}(q^2)$ (upper lines), for $k=0$ and different values of $\alpha_\star$: the solid lines are for $\alpha_\star = 1.5$, the dotted lines are for $\alpha_\star = 1$ and the dashed lines for $\alpha_\star = 2$.}
\renewcommand{\baselinestretch}{1.7}\normalsize
\end{figure}
\bea
\chi_A = - 13 \frac{g_V^2}{16 \pi^2} \qquad , \qquad \chi_{gh} = - \frac{9}{2} \frac{g_V^2}{16 \pi^2}
\label{chifieldspert}
\eea
We have checked that our numerical solution obeys these relations.

\section{Conclusions}

We have computed the effective four quark interaction which results from QCD after integrating out the gluon (and ghost) degrees of freedom.
More precisely, we consider an effective action where also the quark fluctuations with momenta larger than a fermionic infrared cutoff $k_\psi$ are integrated out, but not those with lower momenta.
In Landau gauge the low momentum behaviour of the four quark interaction is dominated by a dressed one gluon exchange whereas contributions from box diagrams and the quark wave function renormalization are less singular.
In fact, the latter can be treated perturbatively since infrared complications are avoided due to the cutoff $k_\psi$ in the quark propagators.
By the use of gauge symmetry (Ward identities) we can express the dominant ``one gluon exchange contribution'' $\tilde{V}(q^2)$ in terms of the gluon and ghost propagators $\overline{G}_A(q^2)$ and $G_{gh}(q^2)$.
These propagators have been computed by an approximate solution of an exact renormalization group equation in the context of the effective average action.

For high enough momenta there is a simple relation between $\tilde{V}(q^2)$ and the effective heavy quark potential $V(q^2)$ in the quenched aproximation ($N_f=0$).
The difference can be computed perturbatively and we have taken into account the one loop contribution.
Using this connection for $\sqrt{q^2} \geq 3 \Lambda_{\overline{\mathrm{MS}}}$ we have optimized our renormalization scheme by employing the existing information on the two loop heavy quark potential in the ${\overline{\mathrm{MS}}}$ scheme.
This also allows us to express all dimensionful quantities in units of the two loop confinement scale $\Lambda_{\overline{\mathrm{MS}}}$.
We should point out that no information on the behaviour of the heavy quark potential for low momenta $\sqrt{q^2} < 3 \Lambda_{\overline{\mathrm{MS}}}$ is required for this optimization.
Our final result for $\tilde{V}$ is the fit in equation (\ref{fit}),
\bea
q^2 \tilde{V}(q^2) = \frac{4}{3} g_V^2(z) + \frac{5 \bar{g}^4}{16\pi^2}
\label{neuneu}
\eea
The full four quark interaction $\hat{V}(q^2)$ (cf. equation (\ref{V})) still needs to be related to $\tilde{V}(q^2)$ by a perturbative calculation, the difference being $\propto \bar{g}^4/q^2$ and subleading for small $q^2$.

Our result (\ref{neuneu}) is expressed in terms of the dimensionless variable $z = q^2 / \Lambda_{\overline{\mathrm{MS}}}^2$.
In contrast to the discusssion of the quenched heavy quark potential in section 6, the effective four quark interaction depends on the scale $k_\psi$.
In fact, the running of the gauge coupling corresponds effectively to a flow with three light quarks, $N_f = 3$, for $k > k_\psi$ and $N_f = 0$ for $k < k_\psi$.
The low momentum range is dominated by the flow for $k < k_\psi$ and the main dependence on $k_\psi$ arises through the scale $\Lambda_{\overline{\mathrm{MS}}}$ in the fit (\ref{fit}).
In fact, $\Lambda_{\overline{\mathrm{MS}}}$ increases with $k_\psi$ due to the faster running of the gauge coupling for $N_f = 0$ as compared to $N_f = 3$.
(To one loop order one finds $\Lambda_{\overline{\mathrm{MS}}} \propto k_\psi^\gamma$, $\gamma = 2/11$.)
For $k_\psi = 1.3 (1.7)$~GeV one obtains numerically $\Lambda_{\overline{\mathrm{MS}}} = 375 (395)$~MeV.
All momentum scales in the figures have to be rescaled correspondingly by a factor $\Lambda_{\overline{\mathrm{MS}}} / 315~{\mathrm{MeV}}$.
With the appropriate choice of $\Lambda_{\overline{\mathrm{MS}}}$ our result (\ref{neuneu}) can now be used as an input for the flow of the effective quark interactions as the fermionic cutoff is lowered
\cite{UliCh}.
For future computations of the flow in the effective quark model an important check of the approximations consists in monitoring if the dependence on $k_\psi$ cancels out of the final results.

We have also made an attempt to use our information on the effective four quark interaction $\tilde{V}(q^2)$ for a computation of the effective heavy quark potential $V(q^2)$ in the quenched approximation.
This involves an additional important assumption, namely that the potential is dominated by a dressed one gluon exchange (in Landau gauge) also for $\sqrt{q^2} < 3 \Lambda_{\overline{\mathrm{MS}}}$.
In other words, we suggest that the difference between $V(q^2)$ and $\tilde{V}(q^2)$ remains subleading for small $q^2$.
Even though suggested strongly by continuity arguments, this remains an assumption as long as no detailed calculations are available.
In the real world the three light quarks contribute to the potential.
(There is no infrared cutoff $k_\psi$ for these quarks.)
In order to adapt our quenched results to this situation one has to choose the relevant scale $\Lambda_{\overline{\mathrm{MS}}}$ appropriatly so that at least the high momentum fluctuations of the quarks are effectively included (see section 6).
Comparing our results to observation (coded in phenomenological potentials) gives then very satisfactory results:
We find an effective potential which interpolates smoothly between the two-loop perturbative potential for high momenta and phenomenologically successful potentials for the range $300~{\mathrm{MeV}} \klgl \sqrt{q^2} \klgl 800~{\mathrm{MeV}}$ which is relevant for quarkonia.
This strongly supports our assumption that $V(q^2)$ is indeed dominated by a dressed one gluon exchange.
If this picture can be substantiated and extends to the straightforward inclusion of light quark fluctuations it will permit to relate the perturbative scale $\Lambda_{\overline{\mathrm{MS}}}$ to quarkonium physics and therefore ultimately compute the fine structure constant at the $Z$-mass $\alpha_s(M_Z)$ in the ${\overline{\mathrm{MS}}}$ scheme.
Although the error estimate is difficult and probably not competitive with a corresponding program in lattice QCD 
\cite{HQPot}
such an analytical link between perturbative and nonperturbative QCD would be of great value.

\end{document}